\documentclass[twocolumn,trackchanges]{aastex631}

\usepackage{CJK}
\usepackage[utf8]{inputenc}
\usepackage{graphicx}
\usepackage{tabularx}
\usepackage{amsmath}
\usepackage{soul}

\newcommand{\degree}{\mbox{$^\circ$}}
\newcommand{\msun}{\mbox{M$_{\odot}$}}

\newcommand{\kms}{\mbox{$\rm{km}\,s^{-1}$}}

\DeclareMathAlphabet{\mathsc}{OT1}{cmr}{m}{sc}
\def\testbx{bx}%
\DeclareRobustCommand{\ion}[2]{%
\relax\ifmmode
\ifx\testbx\f@series
{\mathbf{#1\,\mathsc{#2}}}\else
{\it{#1\,\mathsc{#2}}}\fi
\else\textup{#1\,{\mdseries\textsc{#2}}}%
\fi}
\newcommand{\Ha} {\mbox{H$\alpha$}\,}
\newcommand{\Hb} {\mbox{H$\beta$}\,}

\newcommand{\Nai} {\ion{Na}{i}\,}
\newcommand{\Feii} {\ion{Fe}{ii}\,}

\newcommand{\Hei} {\ion{He}{i}\,}

\newcommand{\Oi} {[\ion{O}{i}]\,}

\received{May 5, 2026}
\revised{May XX, 2026}
\accepted{May 23, 2026}

\begin{document}

\title{The transitional Type Ibn/IIn SN 2022pda, with pre-explosion outbursts and a double-peaked light curve}

\author[0000-0002-7714-493X]{Y.-Z. Cai}
\altaffiliation{yongzhi.cai@inaf.it}
\affiliation{International Centre of Supernovae (ICESUN), Yunnan Key Laboratory of Supernova Research, Yunnan Observatories, Chinese Academy of Sciences (CAS), Kunming, 650216, China}
\affiliation{INAF - Osservatorio Astronomico di Padova, Vicolo dell'Osservatorio 5, 35122 Padova, Italy}

\author[0000-0002-7259-4624]{A. Pastorello}
\affiliation{INAF - Osservatorio Astronomico di Padova, Vicolo dell'Osservatorio 5, 35122 Padova, Italy}

\author[0009-0003-4594-3715]{R. Chiba}
\affiliation{Graduate Institute for Advanced Studies, SOKENDAI, 2-21-1 Osawa, Mitaka, Tokyo 181-8588, Japan}
\affiliation{National Astronomical Observatory of Japan, National Institutes of Natural Sciences, 2-21-1 Osawa, Mitaka, Tokyo 181-8588, Japan}

\author[0000-0003-1169-1954]{T. J. Moriya}
\affiliation{National Astronomical Observatory of Japan, National Institutes of Natural Sciences, 2-21-1 Osawa, Mitaka, Tokyo 181-8588, Japan}
\affiliation{Graduate Institute for Advanced Studies, SOKENDAI, 2-21-1 Osawa, Mitaka, Tokyo 181-8588, Japan}
\affiliation{School of Physics and Astronomy, Monash University, Clayton, Victoria 3800, Australia}

\author[0000-0003-4254-2724]{A. Reguitti}
\affiliation{INAF - Osservatorio Astronomico di Padova, Vicolo dell'Osservatorio 5, 35122 Padova, Italy}
\affiliation{INAF - Osservatorio Astronomico di Brera, Via Bianchi 46, 23807 Merate (LC), Italy}

\author[0000-0003-3433-1492]{L. Tartaglia}
\affiliation{INAF - Osservatorio Astronomico d'Abruzzo, via Mentore Maggini, s.n.c. I-64100 Teramo}

\author[0000-0001-5221-0243]{S. Moran}
\affiliation{Department of Physics and Astronomy, University of Turku, FI-20014 Turku, Finland}
\affiliation{School of Physics and Astronomy, University of Leicester, University Road, Leicester LE1 7RH, UK}

\author[0000-0001-6278-1576]{S. Campana}
\affiliation{INAF - Osservatorio Astronomico di Brera, Via Bianchi 46, 23807 Merate (LC), Italy}

\author[0000-0002-0025-0179]{Z.-Y. Wang}
\affiliation{School of Physics and Astronomy, Beijing Normal University, Beijing 100875, China}
\affiliation{Department of Physics, Faculty of Arts and Sciences, Beijing Normal University, Zhuhai 519087, China}

\author[0009-0009-8633-8582]{J.-W. Zhao}
\affiliation{South-Western Institute for Astronomy Research, Yunnan University, Kunming 650500, China}
\affiliation{Yunnan Key Laboratory of Survey Science, Yunnan University, Kunming, Yunnan 650500, China}

\author[0000-0003-0227-3451]{J.~P. Anderson}
\affiliation{European Southern Observatory, Alonso de C\'{o}rdova 3107, Vitacura, Casilla 19001, Santiago, Chile}

\author[0000-0002-3256-0016]{S. Benetti}
\affiliation{INAF - Osservatorio Astronomico di Padova, Vicolo dell'Osservatorio 5, 35122 Padova, Italy}

\author[0000-0003-1325-6235]{S. J. Brennan}
\affiliation{Max Planck Institute for Extraterrestrial Physics, Max-Planck-Gesellschaft, Giessenbachstra{\ss}e 1, Garching, 85748}

\author[0000-0001-5008-8619]{E. Cappellaro}
\affiliation{INAF - Osservatorio Astronomico di Padova, Vicolo dell'Osservatorio 5, 35122 Padova, Italy}

\author[0000-0001-6965-7789]{K. C. Chambers}
\affiliation{Institute for Astronomy, University of Hawaii, 2680 Woodlawn Drive, Honolulu HI 96822, USA}

\author[0000-0002-1066-6098]{T.-W. Chen}
\affiliation{Graduate Institute of Astronomy, National Central University, 300 Jhongda Road, 32001 Jhongli, Taiwan}

\author{Z.-H. Chen}
\affiliation{Physics Department, Tsinghua University, Beijing, 100084, P.R. China}

\author{T. de Boer}
\affiliation{Institute for Astronomy, University of Hawai'i, 2680 Woodlawn Drive, Honolulu, HI 96822, USA}

\author[0000-0002-7937-6371]{Y.-Z.~Dong}
\affiliation{Center for Astrophysics | Harvard \& Smithsonian, Cambridge, MA 02138, USA}
\affiliation{The NSF AI Institute for Artificial Intelligence and Fundamental Interactions, USA}

\author[0000-0002-1823-3860]{J.~Duarte}
\affiliation{CENTRA, Instituto Superior T\'ecnico, Universidade de Lisboa, Av. Rovisco Pais 1, 1049-001 Lisboa, Portugal}

\author[0000-0002-1381-9125]{N. Elias-Rosa}
\affiliation{INAF - Osservatorio Astronomico di Padova, Vicolo dell'Osservatorio 5, 35122 Padova, Italy}
\affiliation{Institute of Space Sciences (ICE, CSIC), Campus UAB, Carrer de Can Magrans, s/n, E-08193 Barcelona, Spain}

\author[0000-0003-2191-1674]{M. Fraser}
\affiliation{School of Physics, O'Brien Centre for Science North, University College Dublin, Belfield, Dublin 4, Ireland}

\author[0009-0007-7959-9288]{W.-P. Gan}
\affiliation{Nanjing Hopes Technology Co., Ltd. Nanjing, 210000, China}

\author[0000-0003-1015-5367]{H. Gao}
\affiliation{Institute for Astronomy, University of Hawai'i, 2680 Woodlawn Drive, Honolulu, HI 96822, USA}

\author[0000-0002-1650-1518]{M. Gromadzki}
\affiliation{Astronomical Observatory, University of Warsaw, Al. Ujazdowskie 4, 00-478 Warszawa, Poland}

\author[0000-0002-0832-2974]{G. Hosseinzadeh}
\affiliation{Steward Observatory, University of Arizona, 933 North Cherry Avenue, Tucson, AZ 85721-0065, USA}

\author[0000-0003-4253-656X]{D. A. Howell}
\affiliation{Las Cumbres Observatory, 6740 Cortona Dr. Suite 102, Goleta, CA, 93117, USA}
\affiliation{Department of Physics, University of California, Santa Barbara, Santa Barbara, CA, 93106, USA}

\author[0000-0002-3968-4409]{C. Inserra}
\affiliation{Cardiff Hub for Astrophysics Research and Technology, School of Physics \& Astronomy, Cardiff University, Queens Buildings, The Parade, Cardiff, CF24 3AA, UK}

\author[0000-0002-5477-0217]{T. Kangas}
\affiliation{Finnish Centre for Astronomy with ESO (FINCA), University of Turku, Vesilinnantie 5, Quantum 20014 Turku, Finland}
\affiliation{Tuorla  Observatory, Department of Physics and Astronomy, University of Turku, 20014Turku, Finland}

\author[0000-0001-8257-3512]{E. Kankare}
\affiliation{Department of Physics and Astronomy, University of Turku, FI-20014 Turku, Finland}

\author[0000-0003-0955-9102]{T.~Kravtsov}
\affiliation{Department of Physics and Astronomy, University of Turku, FI-20014 Turku, Finland}
\affiliation{Finnish Centre for Astronomy with ESO (FINCA), University of Turku, Vesilinnantie 5, Quantum 20014 Turku, Finland}

\author[0009-0003-3758-0598]{L.-P. Li}
\affiliation{International Centre of Supernovae (ICESUN), Yunnan Key Laboratory of Supernova Research, Yunnan Observatories, Chinese Academy of Sciences (CAS), Kunming, 650216, China}

\author[0000-0002-7272-5129]{C.-C. Lin}
\affiliation{Institute for Astronomy, University of Hawai'i, 2680 Woodlawn Drive, Honolulu, HI 96822, USA}

\author[0000-0002-9438-3617]{T. B. Lowe}
\affiliation{Institute for Astronomy, University of Hawai'i, 2680 Woodlawn Drive, Honolulu, HI 96822, USA}

\author[0000-0002-3664-8082]{P.~Lundqvist}
\affiliation{The Oskar Klein Centre, Department of Astronomy, Stockholm University, AlbaNova, SE-10691 Stockholm, Sweden}

\author[0000-0002-7965-2815]{E. A. Magnier}
\affiliation{Institute for Astronomy, University of Hawai'i, 2680 Woodlawn Drive, Honolulu, HI 96822, USA}

\author[0000-0002-8111-4581]{K.~Matilainen}
\affiliation{Department of Physics and Astronomy, University of Turku, FI-20014 Turku, Finland}

\author[0000-0001-6876-8284]{P. A. Mazzali}
\affiliation{Astrophysics Research Institute, Liverpool John Moores University, Liverpool Science Park, 146 Brownlow Hill, Liverpool L3 5RF, UK}
\affiliation{Max-Planck-Institut f\"ur Astrophysik, Karl-Schwarzschild Str. 1, D-85741 Garching, Germany}

\author[0000-0001-5807-7893]{C. McCully}
\affiliation{Las Cumbres Observatory, 6740 Cortona Dr. Suite 102, Goleta, CA, 93117, USA}

\author[0009-0003-8803-8643]{P. Minguez}
\affiliation{Research Corporation of the University of Hawai'i, 2680 Woodlawn Drive, Honolulu, HI 96822, USA}

\author[0000-0003-3939-7167]{T. E. M\"{u}ller-Bravo}
\affiliation{School of Physics, Trinity College Dublin, The University of Dublin, Dublin 2, Ireland}
\affiliation{Instituto de Ciencias Exactas y Naturales (ICEN), Universidad Arturo Prat, Chile}

\author[0000-0001-9570-0584]{M.~Newsome}
\affiliation{Las Cumbres Observatory, 6740 Cortona Dr. Suite 102, Goleta, CA, 93117, USA}
\affiliation{Department of Physics, University of California, Santa Barbara, Santa Barbara, CA, 93106, USA}
\affiliation{Department of Astronomy, The University of Texas at Austin, 2515 Speedway, Stop C1400, Austin, TX 78712, USA}

\author[0000-0003-0209-9246]{E. Padilla Gonzalez}
\affiliation{Las Cumbres Observatory, 6740 Cortona Dr. Suite 102, Goleta, CA, 93117, USA}
\affiliation{Department of Physics, University of California, Santa Barbara, Santa Barbara, CA, 93106, USA}

\author[0000-0002-7472-1279]{C. Pellegrino}
\affiliation{Las Cumbres Observatory, 6740 Cortona Dr. Suite 102, Goleta, CA, 93117, USA}
\affiliation{Department of Physics, University of California, Santa Barbara, Santa Barbara, CA, 93106, USA}

\author[0000-0002-8041-8559]{P. J. Pessi}
\affiliation{Astrophysics Division, National Centre for Nuclear Research, Pasteura 7, 02-093 Warsaw, Poland}

\author{T. Petrushevska}
\affiliation{Center for Astrophysics and Cosmology, University of Nova Gorica, Vipavska 11c, 5270 Ajdov\v{s}\v{c}ina, Slovenia}

\author[0000-0003-0006-0188]{G. Pignata}
\affiliation{Instituto de Alta Investigaci\`{o}n, Universidad de Tarapac\`{a}, Casilla 7D, Arica, Chile}

\author[0009-0002-2952-7431]{R. P. Santos}
\affiliation{CENTRA, Instituto Superior T\'ecnico, Universidade de Lisboa, Av. Rovisco Pais 1, 1049-001 Lisboa, Portugal}
\affiliation{European Southern Observatory, Alonso de C\'{o}rdova 3107, Vitacura, Casilla 19001, Santiago, Chile}

\author[0000-0001-6797-1889]{S. Schulze}
\affiliation{Center for Interdisciplinary Exploration and Research in Astrophysics (CIERA), Northwestern
University, Evanston, IL, USA}
\affiliation{The Oskar Klein Centre, Department of Physics, Stockholm
University, AlbaNova University Center, Stockholm, Sweden}

\author[0000-0002-8229-1731]{S. J. Smartt}
\affiliation{Department of Physics, University of Oxford, Denys Wilkinson Building, Keble Road, Oxford OX1 3RH, UK}
\affiliation{Astrophysics Research Centre, School of Mathematics and Physics, Queen’s University Belfast, Belfast BT7 1NN, UK}

\author[0000-0001-8605-5608]{I. A. Smith}
\affiliation{Institute for Astronomy, University of Hawai'i, 34 Ohia Ku St., Pukalani, HI 96768-8288, USA}

\author[0000-0001-9535-3199]{K. W. Smith}
\affiliation{Department of Physics, University of Oxford, Denys Wilkinson Building, Keble Road, Oxford OX1 3RH, UK}
\affiliation{Astrophysics Research Centre, School of Mathematics and Physics, Queen’s University Belfast, Belfast BT7 1NN, UK}

\author[0000-0003-1546-6615]{J. Sollerman}
\affiliation{The Oskar Klein Centre, Department of Astronomy, Stockholm University, AlbaNova, SE-10691 Stockholm, Sweden }

\author[0000-0003-4524-6883]{S. Srivastav}
\affiliation{Astrophysics sub-Department, Department of Physics, University of Oxford, Keble Road, Oxford, OX1 3RH, UK}
\affiliation{Astrophysics Research Centre, School of Mathematics and Physics, Queen’s University Belfast, Belfast BT7 1NN, UK}

\author[0000-0002-5571-1833]{M.~D.~Stritzinger}
\affiliation{Department of Physics and Astronomy, Aarhus University, Ny Munkegade 120, DK-8000 Aarhus C, Denmark}

\author[0000-0003-0794-5982]{G.~Terreran}
\affiliation{Adler Planetarium, 1300 S. DuSable Lake Shore Drive, Chicago, IL 60605, USA}

\author[0000-0002-3334-4585]{G. Valerin}
\affiliation{INAF - Osservatorio Astronomico di Padova, Vicolo dell'Osservatorio 5, 35122 Padova, Italy}

\author[0000-0002-1341-0952]{R. Wainscoat}
\affiliation{Institute for Astronomy, University of Hawai'i, 2680 Woodlawn Drive, Honolulu, HI 96822, USA}

\author[0000-0001-7867-9912]{S.-Q. Wang}
\affiliation{Guangxi Key Laboratory for Relativistic Astrophysics, School of Physical Science and Technology, Guangxi University, Nanning 530004, China}

\author{D. R. Young}
\affiliation{Astrophysics Research Centre, School of Mathematics and Physics, Queen’s University Belfast, Belfast BT7 1NN, UK}

\author[0000-0002-1296-6887]{L.~Galbany}
\affiliation{Institute of Space Sciences (ICE, CSIC), Campus UAB, Carrer de Can Magrans, s/n, E-08193 Barcelona, Spain}
\affiliation{Institut d’Estudis Espacials de Catalunya (IEEC), E-08034 Barcelona, Spain}


\author{Z. Li}
\affiliation{International Centre of Supernovae (ICESUN), Yunnan Key Laboratory of Supernova Research, Yunnan Observatories, Chinese Academy of Sciences (CAS), Kunming, 650216, China}

\author[0000-0003-1450-0869]{I. Salmaso}
\affiliation{INAF - Osservatorio Astronomico di Capodimonte, Salita Moiariello 16, 80131 Napoli, Italy}
\affiliation{INAF - Osservatorio Astronomico di Padova, Vicolo dell'Osservatorio 5, 35122 Padova, Italy}

\author[0000-0001-6773-7830]{S. Zha}
\affiliation{International Centre of Supernovae (ICESUN), Yunnan Key Laboratory of Supernova Research, Yunnan Observatories, Chinese Academy of Sciences (CAS), Kunming, 650216, China}

\author{J.-M. Bai}
\affiliation{International Centre of Supernovae (ICESUN), Yunnan Key Laboratory of Supernova Research, Yunnan Observatories, Chinese Academy of Sciences (CAS), Kunming, 650216, China}

\author[0000-0002-3231-1167]{B. Wang}
\affiliation{International Centre of Supernovae (ICESUN), Yunnan Key Laboratory of Supernova Research, Yunnan Observatories, Chinese Academy of Sciences (CAS), Kunming, 650216, China}

\author[0000-0002-7334-2357]{X.-F. Wang}
\altaffiliation{wang\_xf@mail.tsinghua.edu.cn}
\affiliation{Physics Department, Tsinghua University, Beijing, 100084, P.R. China}
\affiliation{Purple mountain observatory, Chinese Academy of Sciences, Nanjing, 210023, P.R. China}

\author[0000-0002-8296-2590]{J.-J. Zhang}
\altaffiliation{jujia@ynao.ac.cn}
\affiliation{International Centre of Supernovae (ICESUN), Yunnan Key Laboratory of Supernova Research, Yunnan Observatories, Chinese Academy of Sciences (CAS), Kunming, 650216, China}

\begin{abstract}
We report the results of a photometric and spectroscopic follow-up campaign of the unusual interacting supernova (SN) 2022pda. Precursor variability lasting $\sim 100$ days is observed before the explosion. The SN light curve has a double peak shape. It reached a first maximum of $M_{\rm{r}} = -19.6 \pm 0.2$\,mag, followed by an initial two-month decline and a second, broad peak lasting about six months.
The early spectra show a blue continuum with dominant H and He emission lines. A high-resolution pre-maximum spectrum shows that the profile of the \Hei~$\lambda$5876 line consists of a moderately narrow ($\sim 1900$ \kms) P~Cygni absorption superposed on a broader ($\sim 3300$ \kms) component.
In the blue region, several spectral features are identified, including C {\sc iii}/N {\sc iii}/O {\sc ii} blends. Two broad bumps at 4600--5200 \AA, 6400--6800~\AA\ regions reveal a complex profile, which are likely due to blends of H, He, and other emission lines.
Late-time spectra are still dominated by prominent and broad H and He lines in emission.
Shock-driven model fits to the bolometric light curve suggest that the SN is powered by interaction with a massive CSM with enhanced mass loss rates $\sim 5$ \msun yr$^{-1}$, expelled during two events occurred $\sim 1$ and $\sim 0.2$ years before the explosion.
The overall SN evolution indicates that SN\,2022pda is a transitional event between a H-rich SN IIn (SN\,2009ip-like) and a He-rich SN Ibn. Our findings suggest that the progenitor was likely a Luminous Blue Variable transitioning towards a Wolf--Rayet stage.
\end{abstract}


\section{Introduction}
\label{sec:intro}
Interacting supernovae (SNe) are explosive transients showing observational signs of shock interaction between their ejecta and pre-existing circumstellar material \citep[CSM; see e.g.,][]{Blinnikov2017hsn..book..843B,Smith2017hsn..book..403S, Fraser2020RSOS....700467F,Dessart2024arXiv240504259D,Gangopadhyay2024arXiv241104107G}. Although a fraction of interacting SNe are thermonuclear explosions of white dwarfs within an H-rich CSM \citep[Type Ia-CSM SNe;][]{Hamuy2003Natur.424..651H,Dilday2012Sci...337..942D,Silverman2013ApJS..207....3S}, in most cases they are core-collapse (CC) SNe embedded in H-rich \citep[Type IIn;][]{Schlegel1990MNRAS.244..269S,Filippenko1997ARA&A..35..309F}, or He-rich/H-poor CSM \citep[Type Ibn; ][]{Pastorello2008MNRAS.389..113P}. For an increasing number of observed SNe, strong evidence suggests that shock interaction seems an ubiquitous phenomenon in stripped envelope (SE) SNe \citep{Niblett2025ApJ...994..259N}.

Type IIn SNe show narrow H lines characterised by full width at half maximum (FWHM) velocities ranging from a few tens to a few hundreds \kms~\citep[e.g.,][]{Kiewe2012ApJ...744...10K, Taddia2013A&A...555A..10T,Salmaso2025A&A...695A..29S}. The radiation produced in the shock interaction between the SN ejecta and the CSM is the main source of powering the luminosity of SNe IIn, and is responsible for the diversity observed in their spectro-photometric properties \citep[e.g., ][]{Taddia2013A&A...555A..10T,Moriya2013MNRAS.435.1520M,Nyholm2020A&A...637A..73N,Ransome2025ApJ...987...13R,Hiramatsu2024arXiv241107287H}. The observed variety suggests that SNe IIn are characterized by a wide range of CSM parameters (expansion velocity, geometry, density profile, distance from the progenitor), hence different progenitor's mass-loss channels and pre-eruptions \citep[][]{Moriya2014MNRAS.439.2917M}. For instance, the multi-peaked light curves of some SNe IIn\footnote{E.g., iPTF13z \citep{Nyholm2017A&A...605A...6N}, SNe\,2005ip \citep{Stritzinger2012ApJ...756..173S}, 2006jd \citep{Stritzinger2012ApJ...756..173S}, 2009ip \citep{Pastorello2013ApJ...767....1P,Fraser2013MNRAS.433.1312F}, 2013gc \citep{Reguitti2019MNRAS.482.2750R}, 2021qqp \citep{Hiramatsu2024ApJ...964..181H}, and 2023zkd \citep{Gagliano2025ApJ...989..182G}.} are likely caused by ejecta collisions with shell-like gas distributions, most likely produced through eruptive mass-loss events rather than steady winds.
 
The spectra of Type Ibn SNe show narrow lines of He~I (with FWHM velocities ranging from hundreds to $\sim$ 1000-3000 \kms) but have little or no evidence of H \citep[e.g.,][]{Pastorello2016MNRAS.456..853P,Hosseinzadeh2017ApJ...836..158H}, revealing the interaction of SN ejecta with He-rich CSM \citep{Smith2008ApJ...680..568S, Smith2017hsn..book..403S,Maeda2022ApJ...927...25M}. SNe Ibn show a moderate heterogeneity in the photometric evolution \citep[][]{Hosseinzadeh2017ApJ...836..158H}. While a limited number of outliers is known\footnote{OGLE-2012-SN-006 \citep{Pastorello2015MNRAS.449.1941P}, iPTF13beo \citep{Gorbikov2014MNRAS.443..671G}, OGLE-2014-SN-131 \citep{Karamehmetoglu2017A&A...602A..93K}.}, most SNe Ibn have fast-evolving light curves, with a short rise time ($\leq$ 1 week), a rapid post-peak decline, and a high luminosity at maximum, reaching peak absolute magnitudes of $-$20.0 $\leq$ $M_{\rm{R}}$ $\leq$ $-$18.5 mag \citep[e.g.,][]{Pastorello2016MNRAS.456..853P,Hosseinzadeh2017ApJ...836..158H,Wang2024A&A...691A.156W}. Such events are generally considered as stripped envelope (SE) CC explosions of massive stars occurring in a He-rich CSM \citep[e.g.,][]{Pastorello2007Natur.447..829P, Moriya2016ApJ...824..100M,Maeda2022ApJ...927...25M}. However, in some cases, their observed parameters can be comfortably explained as the explosion of much lower-mass He stars in binary systems, whose H-rich envelope was stripped by the interaction with the companion star \citep[e.g.,][]{Sanders2013ApJ...769...39S,Hosseinzadeh2019ApJ...871L...9H, Dessart2022A&A...658A.130D}.

In the past decade, we discovered that a significant fraction of SNe IIn may explode a short time after major stellar outbursts \citep[see e.g.,][]{Pastorello2013ApJ...767....1P, Foley2011ApJ...732...32F, Ofek2013Natur.494...65O,Strotjohann2021ApJ...907...99S,Reguitti2024A&A...686A.231R}. 
However, precursor outbursts were also detected from the inspection of archival images of SNe Ibn \citep[e.g., SN\,2023fyq, see][Elias-Rosa et al., in preparation, and references therein]{Brennan2024A&A...684L..18B,Dong2024ApJ...977..254D}. This is an important step to improve our understanding of SNe Ibn, as studying eruptive mass-loss events from their progenitors allows us to probe the latest evolutionary stages of stripped massive stars before their explosion. 

\section{Host environment} \label{sec:host}

The discovery of SN\,2022pda\footnote{The object is also known with survey designations PS22gdx, ATLAS22bbxi and ZTF22aawxlpc.} was announced by the Pan-STARRS Survey for Transients \citep{Fulton2025MNRAS.542..541F} using the 
the Panoramic Survey Telescope and Rapid Response System 2 \citep[Pan-STARRS2 - PS2;][]{Chambers2016arXiv161205560C} 
on 2022 July 19 (MJD=59779.56), and later confirmed by the Asteroid Terrestrial-impact Last Alert System \citep[ATLAS;][]{Tonry2018ApJ...867..105T,Tonry2018PASP..130f4505T, Smith2020PASP..132h5002S} on 2022 August 23 (MJD 59814.22). The classification was performed by \citeauthor{Fulton2022TNSAN.198....1F} on 2022 September 25 as a Type Ibn SN. 
\citeauthor{Fulton2022TNSAN.198....1F} also noted pre-rise activity lasting 70 days before peak in the Pan-STARRS data. 
Its coordinates are $\alpha = 22^{\mathrm{h}} 04^{\mathrm{m}} 13.895^{\mathrm{s}}$, $\delta = -18\degree 49\arcmin 40.46\arcsec$  [J2000], and it is offset by $15.6\arcsec$ south and $10.4\arcsec$ west from the centre of the likely host galaxy LEDA 135305. In the framework of a standard cosmology ($H_0$=73~\kms~$\mathrm{Mpc}^{-1}$, $\Omega_M$=0.27, $\Omega_{\Lambda}$=0.73), we adopted a kinematic distance $d = 253.8 \pm 17.8$\,Mpc ($\mu = 37.02 \pm 0.15$\,mag) for the galaxy, which has been corrected for the influence of the Virgo cluster, the Great Attractor, and the Shapley supercluster \citep{Mould2000ApJ...529..786M}.
The Galactic reddening towards SN\,2022pda is small, $E(B-V)_{\mathrm{Gal}} = 0.022$\,mag \citep{Schlafly2011ApJ...737..103S}, assuming a reddening law with $R_V$=3.1 \citep{Cardelli1989ApJ...345..245C}. To estimate the additional host galaxy reddening, we measured the equivalent width (EW) of the Na\,{\sc i}\,D ($\lambda\lambda$5890,5896) absorption at the redshift of the host galaxy using the early, moderate-resolution SN spectrum taken with the Very Large Telescope (VLT)/X-Shooter. We obtain Na\,{\sc i} D1 and D2 EWs of 0.33 and 0.25 \AA, i.e., EW (D1+D2)$_{\mathrm{Host}}$ = 0.58 \AA.
Following the Equation 9 in \citealt{Poznanski2012MNRAS.426.1465P} (i.e., log$_{10}$($E(B-V)$)=1.17 $\times$ EW (D1+D2) - 1.85 $\pm$ 0.08), we obtain a host galaxy reddening of $E(B-V)_{\mathrm{Host}}=0.07$\,mag. In order to be consistent with the recalibration of Milky Way extinction in \citet{Schlafly2011ApJ...737..103S}, we further multiplied this value by the re-normalisation factor of 0.86, resulting in $E(B-V)_{\mathrm{Host}}=0.06$\,mag \citep[see][]{Gangopadhyay2025MNRAS.537.2898G, Dastidar2025A&A...694A.260D}. 
Throughout the paper, we thus adopt a total line-of-sight color excess of $E(B-V)_{\mathrm{Total}} = 0.08$\,mag. \\

\section{Photometric observations} \label{sec:photometry}

The multi-wavelength optical (Johnson-Cousins $UBV$, Sloan $ugriz$) and near-infrared (NIR; $JHK$) follow-up of SN\,2022pda started soon after its classification, using the telescopes and the instruments listed in Table \ref{table_setup} (Appendix \ref{subsec:facilities}). In addition to the ground-based observations, imaging in the ultraviolet (UV: $UVW2, UVM2, UVW1$) and optical ($UBV$) domains was obtained  through the {\sl Neil Gehrels Swift Observatory} ({\it Swift}) equipped with the {\sl Ultra-violet Optical Telescope} \citep[UVOT; e.g.][]{Roming2005SSRv..120...95R,Poole2008MNRAS.383..627P,Breeveld2010MNRAS.406.1687B} in the $UVW2, UVM2, UVW1, U, B, V$  bands \citep{Gehrels2004ApJ...611.1005G}.
Simultaneously, {\it Swift} X-ray Telescope \citep[XRT;][]{Burrows2005SSRv..120..165B} observed SN\,2022pda on 5 occasions, starting on Sep 27, 2022.
XRT collected a total exposure of 11.9 ks. SN2022pda is not detected with a $3\,\sigma$ upper limit of $1.8\times 10^{-3}$ cts s$^{-1}$ in the 0.3-10 keV energy band. Assuming a power law spectrum with photon index $\Gamma=2$ and the Galactic absorption ($2.2\times 10^{20}$ cm$^{-2}$, this implies an upper limit on the 0.3-10 keV unabsorbed average flux of $4\times 10^{-14}$ erg s$^{-1}$ cm$^{-2}$. At 253.8 Mpc, this translates to an X--ray luminosity $L_X<4\times 10^{41}$ erg s$^{-1}$. Considering  instead only the first 3.7 ks observation taken a few days before the optical peak, the upper limit is 2.2 times higher.

Finally, one epoch of mid-infrared (MIR) photometry in the $W1$ (3.4 $\mu$m) and $W2$ (4.6 $\mu$m) filters was obtained with the Wide-field Infrared Survey Explorer (WISE) space telescope \citep{Wright2010AJ....140.1868W}. 
The reduction techniques adopted for our photometric dataset are extensively discussed in Sect. \ref{subsec:reduphot} (Appendix \ref{Appendix:PhotSpec}).
Along with the observations obtained through our monitoring programs, photometric data are also provided by the major sky surveys: Zwicky Transient Facility \citep[ZTF;][]{Bellm2019PASP..131a8002B,Graham2019PASP..131g8001G}, ATLAS \citep{Tonry2018PASP..130f4505T,Smith2020PASP..132h5002S}
and Pan-STARRS \citep{Chambers2016arXiv161205560C,Fulton2025MNRAS.542..541F}. 
These survey data are important to track the progenitor star activity before the SN discovery. The final UV, optical, and infrared (IR) magnitudes are available at the CDS.

\begin{figure*}[ht!]
\centering
\includegraphics[width=2\columnwidth]{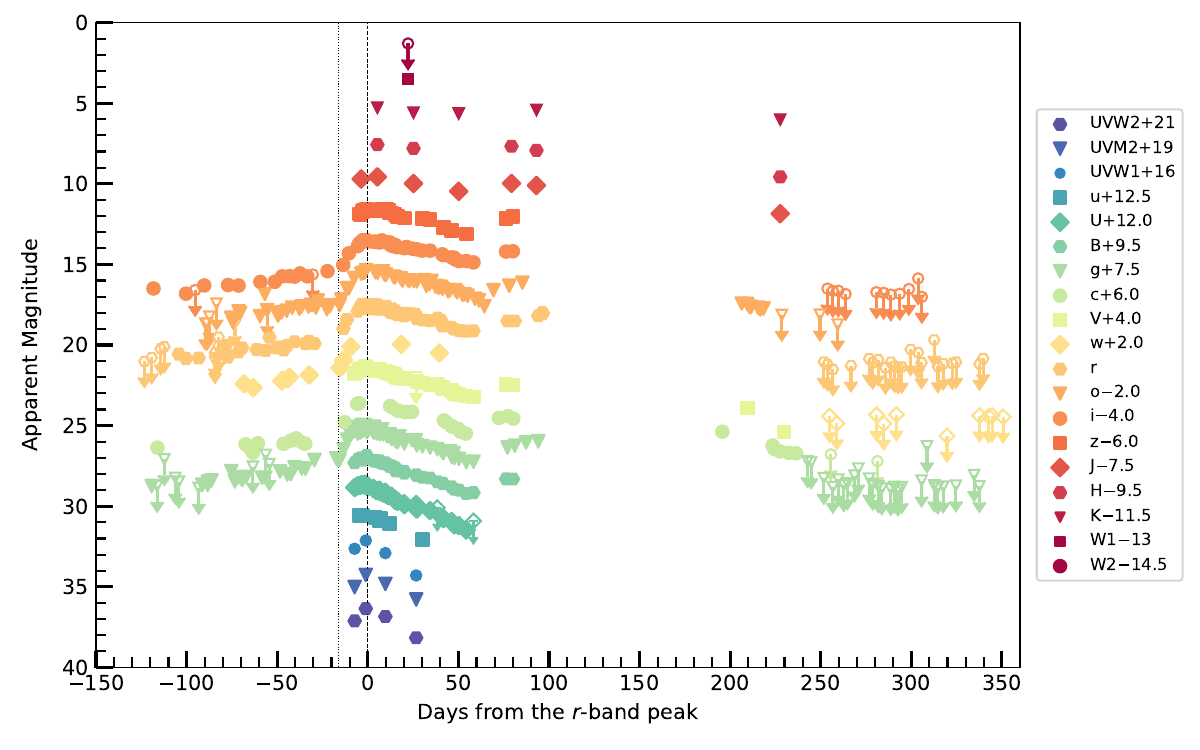}
\caption{Multi-band light curves of SN\,2022pda. The dotted vertical line indicates the estimated explosion epoch, while the dashed vertical line marks the epoch of the $r$-band light curve peak. Upper limits are marked with downward-pointing arrows. Each light curve is shifted by an arbitrary constant indicated in the legend for clarity. The error bars representing the uncertainties in the photometric data, in most of cases, are smaller than the symbols. The phases are referred to the time of $r_{\mathrm{max}}$ ($MJD=59856.7$). 
\label{fig:ApLC}}
\end{figure*}

\subsection{Light curves}

The multi-band (UV-Optical-IR) light curves of SN\,2022pda are shown in Figure \ref{fig:ApLC}. Two main photometric properties are immediately noticeable: 1. signs of photometric variability at the SN location are visible before the SN explosion; 2. the SN light curve clearly shows two humps in all bands.   

The SN\,2022pda site was sampled by the main surveys for $\sim$ 3 years before its discovery. The earliest observation of the field is from Pan-STARRS, and is dated 2019 November 13 ($MJD=58800.222$). At that time, the object was not detected down to an $i$-band limit of 21.5 mag. Then the field was occasionally monitored from 2019 November to 2022 June, but only upper limits were obtained. 
Although the Pan-STARRS discovery was announced on 2022 July 19 at $i=20.28\pm0.12$ mag, an earlier detection was obtained by ZTF on 2022 June 7 ($MJD=59737.436$), with the source being at a magnitude $g=21.24\pm0.32$ mag. 
ATLAS confirmed the SN discovery on 2022 August 23, providing also an earlier detection on 2022 June 10 ($MJD=59740.599$) at $c = 20.36 \pm 0.26$\,mag. 

As shown in Figure \ref{fig:ApLC}, the ZTF, ATLAS, and Pan-STARRS light curves reveal a photometric variability lasting $\sim$100 days (from $MJD=59737$ to $MJD=59839$; with a luminosity rise of $\sim$ 2 mag) prior to  major SN brightening. This 2-mag rise closely resembles those observed in SN\,2009ip-like objects \citep[the so-called `Event~A', see][]{Pastorello2013ApJ...767....1P,Fraser2013MNRAS.433.1312F}. Then, the light curve of SN\,2022pda reached a local minimum on 2022 September 18 ($MJD=59840.3$), which is tentatively assumed to be the SN explosion epoch. 
This minimum is followed by a rapid rise of the light curve ($\sim$2 mag in about two weeks) that reached a first blue peak. The photometric parameters at the SN maximum light were obtained through low-order polynomial fits. The epoch of the $r$-band peak is estimated to be on $MJD = 59856.7\pm0.7$, which will be used as a reference time throughout the paper. The peak magnitude is $r_{\rm{max}} = 17.6 \pm 0.1$\,mag. The light curve shows a fast linear decline after maximum in all bands, with decline rates of 6.61$\pm$0.87, 4.42$\pm$0.35, 2.84$\pm$0.15, and 2.94$\pm$0.55 mag/100 d in the $UVW2$, $B$, $r$, and $z$ filters, respectively. This initial rapid decline is then followed by a slower decline, and a sort of short-duration plateau is observed in the best-sampled red bands. 

After this flattening, the multi-band light curves decline again to a local minimum at $\sim$ $+$60 days, followed by a new brightening. However, the monitoring campaign was suspended at $\sim$ $+$95 days, as the SN was approaching solar conjunction. This late re-brightening likely indicates the onset of an enhanced ejecta-CSM interaction event. After the seasonal gap and the second peak, we obtained a few epochs of late photometric data while the SN magnitude was approaching the instrumental detection thresholds. 
 
\begin{figure*}
\centering
\includegraphics[width=0.8\textwidth]{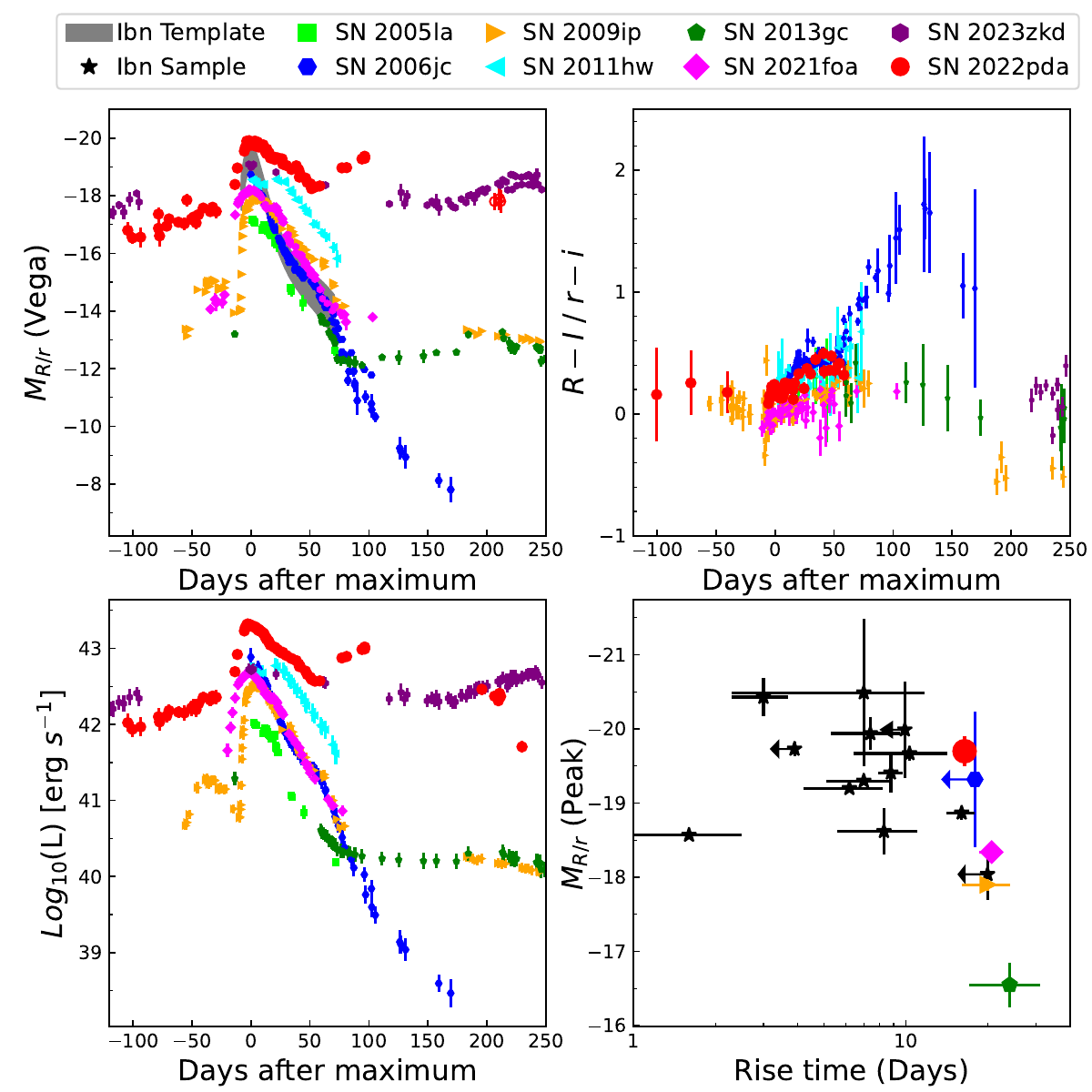}
\caption{Comparisons of $R/r$-band light curves (top-left panel; after seasonal gap, the $o$-band data points are shown in empty circles for completeness.), $R-I/r-i$ color (top-right panel), quasi-bolometric light curves (bottom-left panel), and peak magnitude vs. rise time (bottom-right panel) of SN\,2022pda with those of the transitional Type IIn/Ibn SNe\,2005la, 2011hw, and 2021foa; the Type IIn SNe\,2009ip and 2013gc; and a sample of SNe Ibn. The Type Ibn light curve template is adopted from \citet{Hosseinzadeh2017ApJ...836..158H}. Data sources of comparison objects are from \citet[][]{Matheson2000AJ....119.2303M,Foley2007ApJ...657L.105F,Pastorello2007Natur.447..829P,Pastorello2008MNRAS.389..113P,Pastorello2008MNRAS.389..131P,Pastorello2013ApJ...767....1P,Pastorello2015MNRAS.449.1921P,Pastorello2016MNRAS.456..853P,Smith2010AJ....139.1451S, Smith2012MNRAS.426.1905S,Fraser2013MNRAS.433.1312F,Gorbikov2014MNRAS.443..671G,Shivvers2016MNRAS.461.3057S,Smartt2016ApJ...827L..40S,Hosseinzadeh2017ApJ...836..158H,Reguitti2019MNRAS.482.2750R,Reguitti2022A&A...662L..10R,Wang2020ApJ...900...83W,Farias2024ApJ...977..152F,Gangopadhyay2025MNRAS.537.2898G,Wang2024A&A...691A.156W}. 
\label{fig:LCcomp}}
\end{figure*}

In Figure \ref{fig:LCcomp}, we compare the $R/r-$band absolute light curves (top-left panel), the $R-I/r-i$ color curves (top-right panel), the quasi-bolometric light curves (bottom-left panel), and the peak magnitude vs. rise time parameters (bottom-right panel) of SN\,2022pda with those of the transitional IIn/Ibn SNe\,2005la, 2011hw and 2021foa; the H-rich SNe\,2009ip  and 2013gc; the double-peaked, He-rich Type IIn SN\,2023zkd, and a sample of Type Ibn SNe. The $R/r-$band absolute light curve shows that SN\,2022pda is characterized by a double-hump light curve, resembling those of the Type IIn SN 2013gc, SN\,2023zkd, and some transitional IIn/Ibn SNe. Another remarkable feature in the light curve of SN\,2022pda is its pre-SN variability, which is similar to those seen before some SNe IIn \citep[e.g., SNe\,2009ip, 2010mc, 2013gc, 2015bh, AT\,2016jbu, SN\,2023zkd; see][]{Pastorello2013ApJ...767....1P,Pastorello2018MNRAS.474..197P,Fraser2013MNRAS.433.1312F,Ofek2013Natur.494...65O,Elias-Rosa2016MNRAS.463.3894E,Reguitti2019MNRAS.482.2750R,Brennan2022MNRAS.513.5642B,Gagliano2025ApJ...989..182G}, and - more rarely - in transitional SNe Ibn/IIn \citep[e.g., SN\,2021foa;][]{Reguitti2022A&A...662L..10R,Gangopadhyay2025MNRAS.537.2898G,Farias2024ApJ...977..152F} and in the SN~Ibn SN~2023fyq \citep{Brennan2024A&A...684L..18B,Dong2024ApJ...977..254D}.  Furthermore, the peak absolute magnitude of all SNe in the comparison lies in the range $-$17 to $-$19 mags, while SN\,2022pda (with $M_{\rm{r}}$=$-$19.6$\pm$0.2\,mag) is the brightest event in the sample.

During the pre-SN phase, the $r-i$ color of SN 2022pda shows a modest increase from $\sim$0.2 mag at $\sim$$-100$ days to $\sim$0.3 mag at $\sim$$-70$ days, followed by a gradual blueward evolution to $\sim$0.2 mag at $\sim$$-40$ days and further to $\sim$0.1 mag at $\sim$$-6$ days (see the top-right panel of Figure \ref{fig:LCcomp}). Around maximum light, the color remains relatively stable, fluctuating within $\sim$0.1-0.3 mag from $-5$ to $+15$ days without a clear monotonic trend. After this phase, the $r-i$ color begins an overall redward evolution, increasing from $\sim$0.2 mag to a peak of $\sim$0.5 mag at $\sim$$+40$ days. At later phases, the color becomes bluer again, decreasing to $\sim$0.3 mag by $\sim$$+60$ days, at a time coincident with the re-brightening seen in the light curve.
SN\,2021foa has a similar color evolution, but it is systematically bluer than SN\,2022pda. The Type IIn SN\,2009ip has a well-sampled pre-maximum color curve  with a nearly flat color evolution ($r-i$ $\sim$0 mag), while its color becomes bluer again ($\sim$$-$0.5 mag)  at late-times. SN\,2006jc and SN\,2011hw show a similar color evolution at phases $< +$70 days, both moving towards slightly redder colors. In particular, the $(r-i)$ color of SN\,2011hw reaches $\approx$ 0.7 mag at about $+$70 days. Later, the color of SN\,2006jc increases rapidly from 0.8 mag (at $+$70 days) to 1.7 mag (at $+$130 days), which is likely associated with the formation of dust \citep[][]{DiCarlo2008ApJ...684..471D,Mattila2008MNRAS.389..141M,Anupama2009MNRAS.392..894A}.   
While the Type IIn SN\,2013gc and SN\,2023zkd do not have early observations, their $r-i$ colors show a modest evolution from $-0.1$ mag to 0.4 mag at phases later than $+$50 days. As expected, the comparison shows a large heterogeneity in the color evolution of interacting SNe.

We then computed a quasi-bolometric light curves for SN\,2022pda and the comparison objects, following the prescriptions of \citet[][]{Cai2026A&A...707A.157C}. To proceed with a meaningful comparison between those transients, we constructed quasi-bolometric light curves integrating the fluxes in the optical bands, accounting for the distances and reddening estimates of each object. As shown in the bottom-left panel of Figure \ref{fig:LCcomp}, SN\,2022pda, SN\,2011hw, and SN\,2023zkd show a pronounced double-peak bolometric light curve. SN\,2022pda with $L_{\rm{SN\,2022pda}}\sim 2.1\times10^{43}$ erg s$^{-1}$ is significantly more luminous than the other objects at peak. For reference, for four comparison objects, we determined the following values: $L_{\rm{SN\,2011hw}}\sim 5.2\times10^{42}$ erg~s$^{-1}$, $L_{\rm{SN\,2009ip}}\sim$ $2.9\times10^{42}$ erg s$^{-1}$,  $L_{\rm{SN\,2005la}}\sim 1.1\times10^{42}$ erg s$^{-1}$, and $L_{\rm{SN\,2023zkd}}\sim$ $6\times10^{42}$ erg s$^{-1}$).  

Figure \ref{fig:LCcomp} (bottom-right panel) shows the peak absolute magnitude vs. rise time diagram including SN\,2022pda and other ejecta-CSM interacting comparison objects. This plot shows that SN\,2022pda has an intermediate  peak absolute magnitude ($M_{\rm{r}} \approx$ $-$19.6\,mag) in the context of the Type Ibn SN sample, whereas it lies in the slower tail ($\sim$16 days) of the rise time distribution.  We remark that transitional Type IIn/Ibn events have a wide range of observed peak absolute magnitudes ($-19.6 \leq M_{\rm{R/r}} \leq -16.5$ mag), and in general longer rise times ($16 \leq t \leq 24$ days) than typical SNe Ibn ($t \leq 10$ days).\\

\section{Spectroscopic observation} \label{sec:spectroscopy}

Following the classification of SN\,2022pda, we launched a spectroscopic follow-up campaign using a number of facilities, during which we obtained 19 spectra spanning a period from about $-$10 days to $+$70 days relative to maximum light. Then, spectroscopic monitoring was suspended because the object approached the solar conjunction. Usually, the spectra were reduced following standard prescriptions, but in some cases we used dedicated pipelines. The data reduction steps and information on the spectra of SN\,2022pda are detailed in Appendix \ref{subsec:spectdata}.  
The basic parameters of the instruments used are reported in Table \ref{2022pdaSpecInfo}. 
The resulting spectra of SN\,2022pda are shown in Figure \ref{fig:spectra}.\\

\subsection{Spectral evolution} \label{sec:spectra}

\begin{figure*} 
\centering
\includegraphics[width=0.81\textwidth]{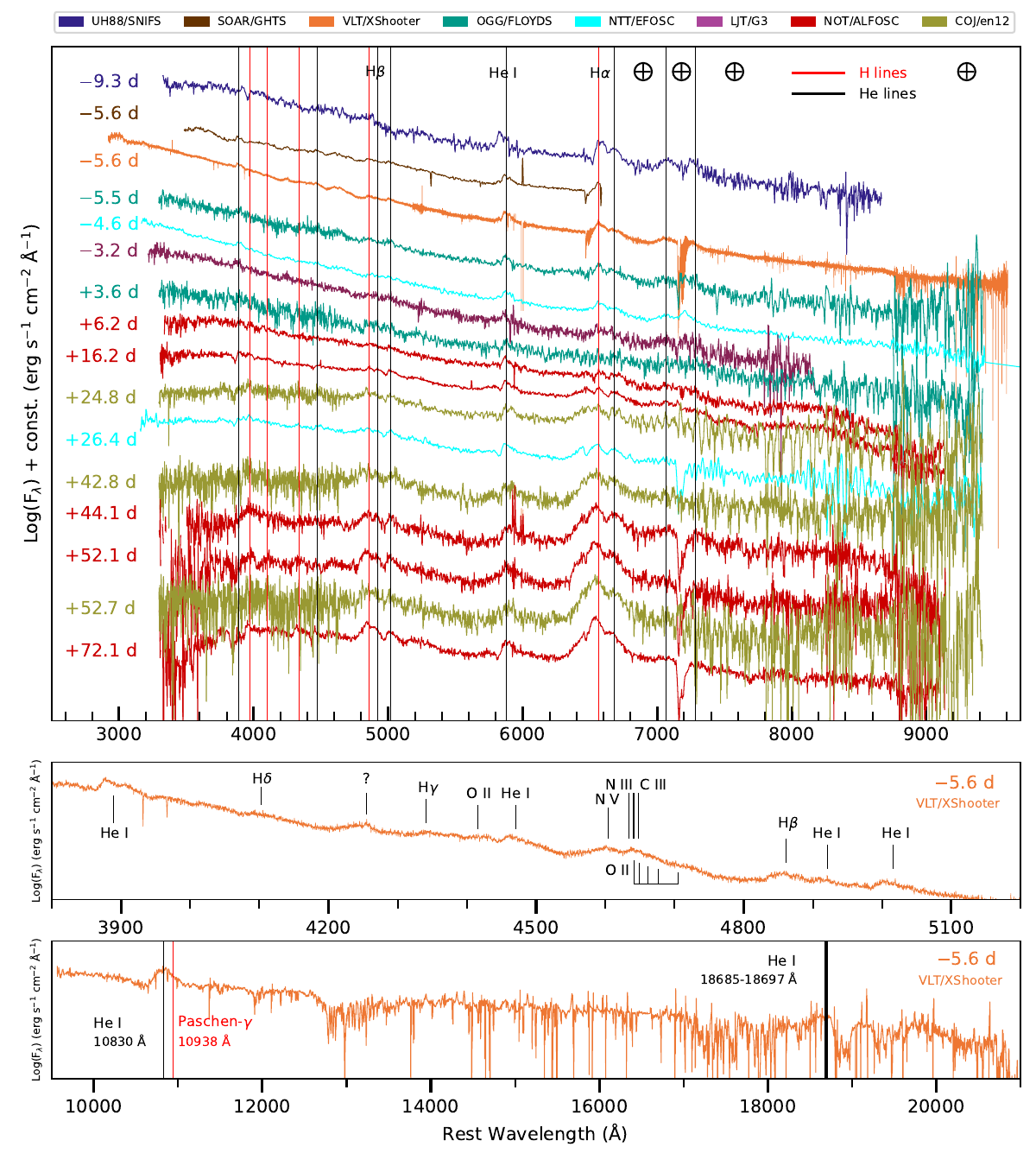}
\caption{Spectroscopic evolution of SN\,2022pda. Top panel: Sequence of optical spectra between phases $-$9.3 d and +72~d. Vertical lines mark the positions of the most prominent Balmer and  \Hei~lines. Middle panel: Identification of the most prominent lines in the blue part (3800--5200 \AA) of the VLT/X-Shooter spectrum obtained on 2022 September 29. Bottom panel: Identification on the $-$5.6 d VLT/X-Shooter NIR spectrum. This spectrum has been binned to 20 \AA~to increase the S/N. Reddening and redshift corrections have been applied in all spectra. The positions of most significant telluric bands are marked with the symbol `$\bigoplus$'. Some lower S/N spectra are not shown in this figure, while they are listed in the Table \ref{2022pdaSpecInfo} (Appendix \ref{subsec:spectdata}).
\label{fig:spectra}}
\end{figure*}

The earliest spectrum shows a blue continuum with a black-body temperature $T_{\rm{BB}}$ $\geq$ 15\,500~K \footnote{This value is considered a lower limit due to the lack of UV spectra, especially since the emission is UV-dominated during this early phase.} (at $-$9.3 days), along with narrow He and prominent \Ha~emission lines, suggesting that it is a transitional Type Ibn/IIn event. 
Four days later (at $-$5.6 days), we obtained a higher resolution optical+NIR spectrum using VLT/X-Shooter. In this spectrum,  emission lines of \Hei~$\lambda\lambda$3889, 4471, 4922, 5016, 5876, 6678, 7065, and 7281 are clearly detected, along with prominent \Ha and \Hb. As highlighted in the middle panel of Figure \ref{fig:spectra}, we identify numerous spectral features in the blue region (3800 \AA~- 5200 \AA) of this spectrum. In particular, the broadest bump at 4600-4700 \AA~is tentatively identified as a blend of C {\sc iii}/N {\sc iii}/O {\sc ii} \citep{Modjaz2009ApJ...702..226M}. Such lines can be interpreted as evidence of CNO-element enrichment in the progenitor's CSM. This suggests enhanced mixing from the inner CNO burning regions to the outer layers. The \Hei $\lambda$5876 line has a clear P~Cygni profile, with an intermediate-width Lorentzian emission component (with a FWHM velocity of about 3300 \kms)~and a blue-shifted Gaussian 
absorption. The position of the blue-shifted absorption minimum indicates that the He-rich material is moving at a velocity of $\sim$ 1900 \kms. This line is likely produced in an unshocked CSM surrounding the SN. 
In the NIR spectrum (bottom panel of Figure \ref{fig:spectra}), the most remarkable bump at 10250 \AA~- 11250 \AA~is possibly due to a blend of \Hei $\lambda$10830 and Paschen-$\gamma$~$\lambda$10941 emission lines. We measured the position of the core of the blueshifted broad P~Cygni absorption of \Hei $\lambda$10830, which is considered as a proxy for the velocity of the SN ejecta. Another shallow bump peaks near 18690 \AA, which can be tentatively identified as \Hei~$\lambda$18685-18697. However, there is no clear evidence that H (e.g., Paschen and Brackett series) lines can be securely identified in this NIR spectrum, likely because of the quite low S/N. 

From $-$5.6 days to $+$6.2 days, the spectra show a minor evolution in terms of feature strengths and line velocities. The H~I and He~I lines are still strong and the pronounced bump visible in the 6400-6800 \AA~region is due to a blend of \Ha and \Hei $\lambda$6678. The velocities derived from deblending the \Hei $\lambda$5876 line are $\sim$1900-2000 \kms~from the wavelength of the P~Cygni minimum, and $\sim$3300-4350 \kms~from the FWHM of the Lorentzian emission component, similar to what measured in the earlier spectra. 
The spectral continuum still remains quite blue with $T_{\rm{BB}}$ in the range 11\,000 K to 18\,000 K. Subsequently, spectra at phases from $+$16.2 to $+$26.4 days show a redder continuum, with $T_{\rm{BB}}$ decreasing to 8\,700$\pm$700 K (at $+$26.4 days). The line velocities increase with a broader emission component at $v_{\mathrm{FWHM}}$ $\approx$~4950 \kms.
\Ha~and the \Hei~$\lambda$6678 blend becomes more prominent, and now shows a clear ``W"-shape profile. Another  prominent feature is visible at 4600-5200~\AA, likely a blend of \Hb, \Hei $\lambda$4922, and \Hei $\lambda$5016. In the last five spectra of SN\,2022pda, obtained from $+$42.8 to $+$72.1 days, all the emission features become more prominent. The velocity of the \Hei\ $\lambda$5876 line further increases, with a~ $v_{\mathrm{FWHM}}$ $\approx$ 5800 \kms~at $+$44.1 day and $v_{\mathrm{FWHM}}$ $\approx$ 6700 \kms~at $+$52.7 day. He I $\lambda$5876 with a narrow P~Cygni profile is detected in the last spectrum ($+$72.1 day), with a velocity of about 2800 \kms, inferred from the position of the minimum of the P~Cygni absorption relative to zero velocity position. In addition, we infer a FWHM velocity of the broadest component of the strong 6400-6800 \AA~bump of $v_{\mathrm{FWHM}}\approx12640$ \kms~in the $+$44.1 d spectrum, which decreases slightly to $v_{\mathrm{FWHM}}\approx10230$ \kms~at $+$72.1~d.

It is worth noting that these late-time features have alternative plausible identifications. For instance, the bump at 4600-5200 \AA~can be a blend of \Hb, \Hei, and \Feii (multiplet~42), while the emission feature at about 5880 \AA~may also be due to the contribution of \Nai D. Finally, the pronounced triple-peaked profile at 6400-6800 \AA~is likely due to high-velocity \Ha~components with a blue and a red emission bumps.
The temperature of the continuum continued to decrease to a minimum $T_{\rm{BB}}$=6\,620$\pm$340~K at $+$52.7 d and subsequently increased to $T_{\rm{BB}}$=7\,120$\pm$100~K at $+$72.1 d. This late-time rise in temperature is consistent with the observed light-curve rebrightening.

\begin{figure*} 
\centering
\includegraphics[width=0.9\textwidth]{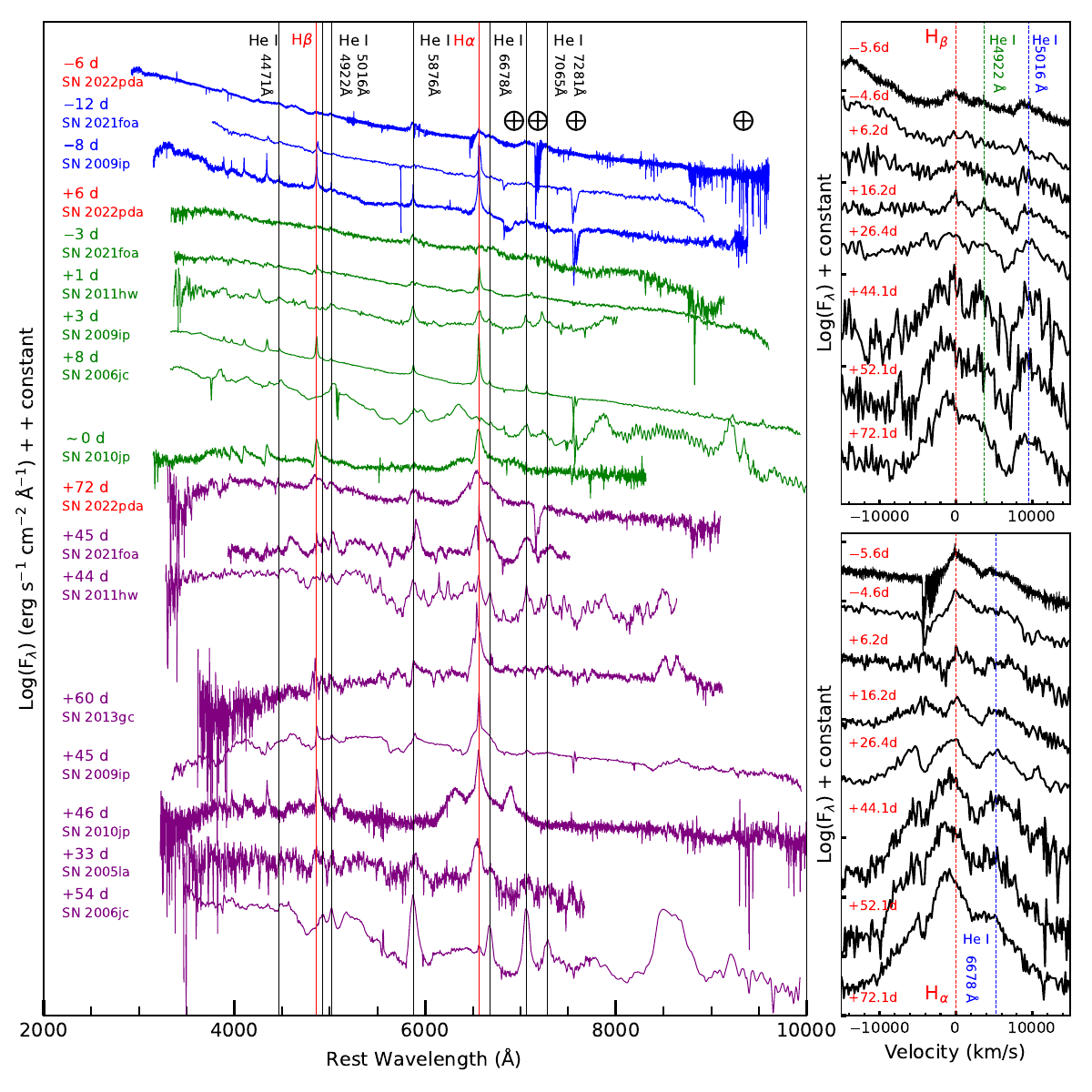}
\caption{Left panel: Comparison of SN\,2022pda spectra with those of reference objects at similar phases: before the light-curve peak (blue), around maximum (green), and at late times (purple). The most relevant lines (\Ha, \Hb, and the main \Hei~lines) are marked with vertical lines. The `$\bigoplus$' symbols mark the principal telluric absorption. Right panel: Close up of the regions of \Ha~and \Hb~of SN\,2022pda spectra in velocity space. \Ha, \Hb, and \Hei~lines are highlighted by vertical lines.   
\label{fig:spectraComp}}
\end{figure*}

In Figure \ref{fig:spectraComp}, we compare the spectra of SN\,2022pda obtained before maximum, around maximum, and at late times with the prototypical Type Ibn SN\,2006jc, the transitional Type IIn/Ibn SNe\,2005la, 2011hw, and 2021foa, 
and the Type IIn SNe\,2009ip, 2010jp, and 2013gc at similar phases. As shown in the left panel of Figure \ref{fig:spectraComp}, all early spectra obtained before and around maximum light show a blue continuum with superposed narrow H and He lines. In this phase, SN\,2022pda shows a blue continuum and narrow \Hei~lines in the spectra, which is similar to those of other SNe Ibn (e.g., SN\,2006jc). SN\,2022pda and other transitioning SNe IIn/Ibn display also H lines, which are  much weaker than in SN\, 2021foa and the H-rich SNe\,2009ip and 2010jp. This comparison highlights the existence of a continuum in the spectral properties from SNe IIn to SNe Ibn, with SN\,2022pda being a bridge connecting these two types. We note that the velocity of the \Hei~$\lambda$5876 line (1900 \kms) in SN\,2022pda is on the higher end, but still within the range of velocities measured in large samples of SNe Ibn at maximum light \citep[see][]{Pastorello2016MNRAS.456..853P, Wang2024A&A...691A.156W}. 

Late-time ($\ge$30 days) spectra of SN\,2022pda and the comparison objects show a major evolution. The continuum becomes redder, the H and He lines are now stronger and broader, while other lines (such as \Oi~$\lambda\lambda$7774-7777 and $\lambda$8446) appear in the redder spectral region. In particular, the NIR \ion{Ca}{ii}~triplet becomes clearly visible in SNe\, 2011hw, 2013gc and 2006jc. We also note that the pronounced bump at 6400-6800 \AA\ (tentatively identified as a triple-peaked \Ha\ in SN\,2022pda) shows a striking similarity with the \Ha~profile of SN\,2010jp. 
SN\,2010jp displays an unprecedented triple-peaked \Ha~line profile, characterized by a narrow central component (FWHM velocity $\geq$800 \kms), two bumps centered at $-$13\,000 \kms~(blue emission) and $+$15\,000 \kms~(red emission), and very broad wings extending from $-$22\,000 to $+$25\,000 \kms~\citep[see Fig. 7 in ][]{Smith2012MNRAS.420.1135S}.
This feature was interpreted by \citet{Smith2012MNRAS.420.1135S} as a bipolar jet-driven explosion, where the central peak arises from ejecta-CSM interaction at mid and low latitudes and  the high-velocity components originate from a non-relativistic bipolar jet. As shown in Figure \ref{fig:spectraComp} (right panel),  the $+$72.1 d spectrum of SN\,2022pda shows a similar \Ha~profile but with lower velocities for the two bumps centered at $-$5\,250 \kms~and $+$4\,000 \kms, and broad wings extending from $-$1\,4000 to $+$1\,3000 \kms. However, the narrow central component of SN\,2022pda  exhibits a relatively higher velocity than that of SN\,2010jp, with an FWHM velocity of $\sim$3650 \kms. 

The right panels of Figure \ref{fig:spectraComp} shows in detail  the evolution of the \Hb\ (and \Hei~$\lambda$4922, 5016; top) and \Ha\ (and \Hei~$\lambda$6678; bottom) line profiles of SN\,2022pda in the velocity space. The \Ha~line shows an asymmetric shape with a two-component profile: a narrow absorption component at about $-$4000 \kms~plus an emission feature at about $-$5000 \kms. As shown in the right top panel of Figure \ref{fig:spectraComp}, the \Hb~profile and its evolution are similar to that of \Ha. 

\section{Discussion and conclusion} \label{sec:discussion}
\subsection{Pre-SN activity} \label{pre-SNactivity}
SN\,2022pda shows a remarkable light curve with a moderately long ($\sim$ 100 days) pre-SN activity and two dominant post-explosion peaks.

A light-curve flattening is visible between 2022 October 29 ($MJD=59881$) and November 8 ($MJD=59891$), with a midpoint at $MJD=59886$, hence approximately 46 days after the explosion. In analogy with similar features observed in other interacting SNe \citep[e.g., SNe\,2009ip, 2013gc, 2018cnf; see][]{Martin2015AJ....149....9M,Reguitti2019MNRAS.482.2750R,Pastorello2019A&A...628A..93P}, this short-duration plateau of SN\,2022pda is likely produced by enhanced shock interaction between the SN ejecta and material expelled during a pre-SN outburst. Adopting a velocity of 4900 \kms~for the SN ejecta (as estimated from the position of the minimum of the broad P~Cygni profile of \Hei~$\lambda$10830, see Sect.\ref{sec:spectra}), they reach the innermost He-rich CSM layers at a distance of about $1.9 \times 10^{15}$ cm.  Assuming the He-rich CSM velocity is about 1900 \kms\ (see Sect.\ref{sec:spectra}), this material was possibly ejected at about 0.32 year ($\sim$119 days) before the explosion, which is roughly in agreement with the timescales of the observed pre-SN eruptive phase. We remark that similar eruptive phases have been observed in other interaction-powered SNe \citep[e.g.,][and references therein]{Strotjohann2021ApJ...907...99S}, and were tentatively explained as due to binary interaction \citep[][]{Smith2014ARA&A..52..487S,Dessart2022A&A...658A.130D,Metzger2022ApJ...932...84M,Tsuna2024ApJ...966...30T}.

\subsection{Light curve modelling} \label{sec:modeling}

The persistent H and He emission lines and the light curve morphology suggest that there is a significant contribution to the SN luminosity from the interaction between the SN ejecta and a dense CSM.
Following this motivation, we model the bolometric light curve of SN 2022pda derived from blackbody fits to the broad-band photometry, assuming that the radiation is fully powered by ejecta--CSM interaction.

We first discuss the effect of extended shock breakout inside CSM.
As noted extensively in the previous literature, the large optical thickness of dense CSM can give rise to longer shock breakout phase (see e.g., \citet{Chevalier2011ApJ...729L...6C}, \citet{Ginzburg2012-av}, as well as \citet{Chiba2024ApJ...973...14C} for double-peaked interacting SNe in particular) than typical core-collapse SNe.
However, as pointed out in \citet{Chiba2024ApJ...973...14C}, it is difficult to reconcile double-peaked light curves with this interpretation when (1) the second peak is comparably bright to the first peak and (2) the separation between the peaks is longer than the rise time of the first peak.
This is because the density requirement for a bright second peak is in conflict with the relatively rapid rise from the breakout.
Since the light curve in Figure~\ref{fig:LCcomp} satisfies both of the condition, we consider the scenario where the shock breakout dominates the first peak to be unlikely.

To reconstruct the CSM density profile, we follow the approach of \citet{Hiramatsu2024ApJ...964..181H}, which is based on the formalism of \citet{Moriya2013MNRAS.435.1520M} that derives the contribution of the interaction to the luminosity under the approximation of a thin shell for the emitting region \citep[also see; e.g.,][]{Chevalier1982ApJ...258..790C,Chevalier1994ApJ...420..268C,Matzner1999ApJ...510..379M,Svirski2012ApJ...759..108S}.
This method has been used to interpret the light curves of SN 2021qqp \citep{Hiramatsu2024ApJ...964..181H} and SN~2023zkd \citep{Gagliano2025ApJ...989..182G}, both interacting SNe with double-peaked light curves and precursor detections.

We assume that the ejecta follows a broken power-law density profile with $\delta = 0$, $n = 10$ \citep[corresponding to the explosion of stars with radiative envelopes;][]{Matzner1999ApJ...510..379M}.
The radiative efficiency $\varepsilon$ is set to a constant value of $\varepsilon = 0.3$ for both the forward and reverse shocks, in line with the observation of interacting SNe \citep[e.g.,][]{Fransson2014ApJ...797..118F}.
Based on the analysis of the He emission lines in Section~\ref{sec:spectra}, we set the CSM velocity to $1900 \ \mathrm{km \ s^{-1}}$.
We assume that the interaction started $0.1 \ \mathrm{d}$ after the explosion, and the initial shell velocity is assumed to be $10000  \ \mathrm{km \ s^{-1}}$, in line with \citet{Gagliano2025ApJ...989..182G}.

\begin{figure*}
    \centering
    \includegraphics[width=1\linewidth]{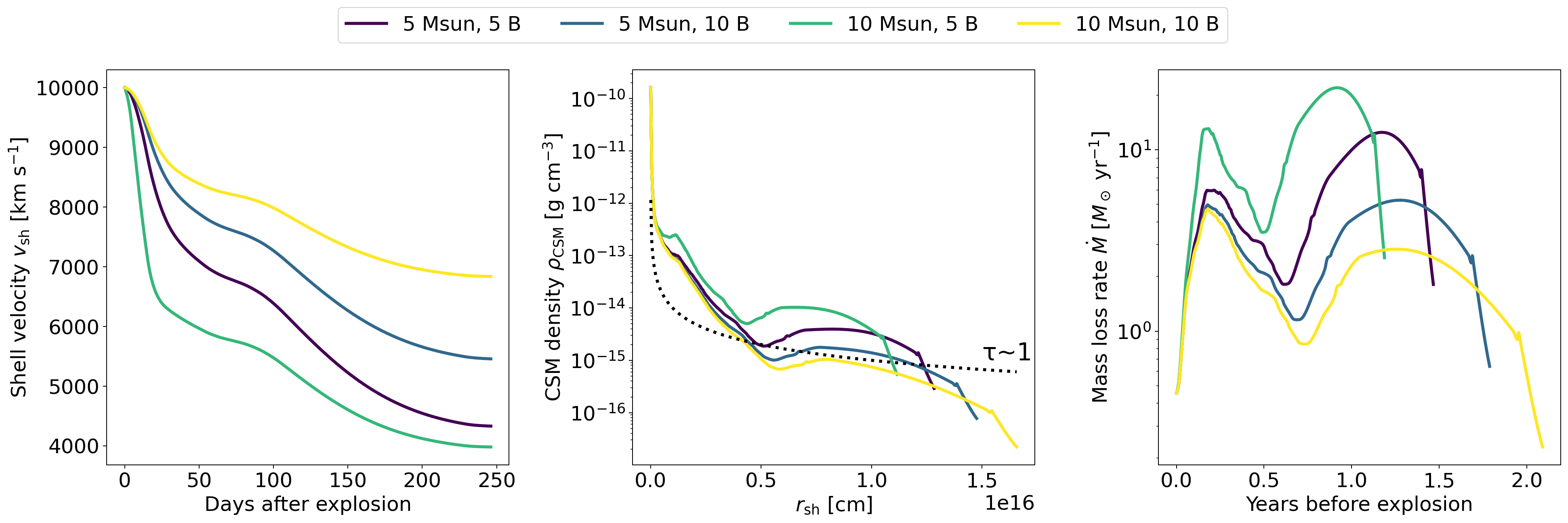}
    \caption{
    CSM interaction models with different assumptions for ejecta masses and explosion energies.
    Left panel: Evolution of the shell velocity.
    Middle panel: CSM density distribution reconstructed from bolometric light curve of SN 2022pda for different ejecta masses and explosion energies. 
    The dotted line shows the rough estimate of CSM optical depth, $\rho = 1 / (\kappa r_\mathrm{sh})$ with a fiducial value of $\kappa = 0.1 \ \mathrm{cm^2 / g}$ representing the scattering opacity of ionised H-poor CSM.
    Right panel: Evolution of pre-explosion mass loss rate under the assumption of constant CSM expansion velocity $v_\mathrm{CSM} = 1900 \ \mathrm{km/s}$.}
    \label{fig:csm_results}
\end{figure*}

We consider different combinations of ejecta masses and explosion energies: $(M_\mathrm{ej}, E_\mathrm{ej}) = (5\, \msun, 5\, \mathrm{B}); (5\, \msun, 10\, \mathrm{B}); (10\, \msun, 5\, \mathrm{B});$ and $(10\, \msun, 10\, \mathrm{B})$.
The evolution of the shell velocity in each model is shown in the left panel of Figure~\ref{fig:csm_results}, while the reconstructed CSM density profile for each combination of explosion parameters is shown in the middle panel of Figure~\ref{fig:csm_results}.
The total CSM masses in each model are $13.5\, \msun,~$8.0\, \msun, $21.3\, \msun$, and $5.5\, \msun$, respectively.
As noted in \citet{Hiramatsu2024ApJ...964..181H}, higher masses of CSM are required for lower explosion energies, in order to extract a larger fraction of kinetic energy from the ejecta.
Heavier CSM mass is required for SN 2022pda than similar, less luminous objects in literature: 2.7\, \msun~for (10\, \msun, 4\, $\mathrm{B}$) in SN 2021qqp \citep{Hiramatsu2024ApJ...964..181H}, 5-6\, \msun~for (10\, \msun, 2\, $\mathrm{B}$) in SN 2023zrk \citep{Gagliano2025ApJ...989..182G}.

We can convert the CSM density profile into pre-explosion mass loss rate through an assumption of the CSM expansion velocity.
Assuming a constant $v_\mathrm{CSM} = 1900 \ \mathrm{km/s}$, the resulting evolution of the mass loss rate is shown in the right panel of Figure~\ref{fig:csm_results}.
In order to power the double-peaked light curve of SN 2022pda, two separate mass loss events are required: both of them with a mass-loss rate of $\sim 5 \ M_\odot \ \mathrm{yr^{-1}}$, but occurred  $\sim 0.2$ years (roughly consistent with the detected pre-SN activity; see discussion in Sect. \ref{pre-SNactivity}) and $\sim 1$ year before the explosion, respectively.

Finally, we discuss the effect of CSM optical thickness.
As pointed out in \citet{Hiramatsu2024ApJ...964..181H}, it is not entirely trivial whether we should expect CSM to be optically thick or thin in this interpretation.
On one hand, a denser CSM is favourable in order for the reprocessing of dissipated kinetic energy to radiation to be effective.
On the other hand, in order to sustain a longer interaction powered phase, CSM should be less dense for the shock breakout to occur earlier.
The more detailed modelling incorporating the effects of diffusion inside CSM and and asphericity is deferred to future works.


\subsection{Spectroscopic implication}

In general, the spectroscopic evolution of SN\,2022pda is similar to those of typical SNe Ibn in the shape of pseudo-continuum, the line identification, the line profiles and velocities (see details in Sect. \ref{sec:spectroscopy}). However, the main difference between normal SNe Ibn and SN\,2022pda is the existence of prominent H emission lines. This feature suggests that SN\,2022pda is a new case of transitioning SN Ibn/IIn, occasionally observed in the past \citep[e.g., SNe 2005la, 2011hw, 2021foa; see][]{Pastorello2008MNRAS.389..131P,Smith2012MNRAS.426.1905S,Pastorello2015MNRAS.449.1921P,Reguitti2022A&A...662L..10R,Farias2024ApJ...977..152F,Gangopadhyay2025MNRAS.537.2898G}. However, the discovery of these transitional interacting SNe suggests a continuum in the properties between at least some SNe IIn and SNe Ibn. The intensity of the H lines gradually decreases from the H-dominated SNe IIn, through SN\,2009ip-like (with strong H plus weak He), SN\,2021foa (with strong H and He), to SNe Ibn (only showing narrow He~I), with SN\,2022pda sitting somewhere between them. In this context, SN\,2022pda and other  SN Ibn/IIn events are proposed to result from the explosion of massive stars that were transiting from the LBV to the WR stages \citep[see Sect. \ref{subsec:progenitor}, and][]{Pastorello2008MNRAS.389..131P,Smith2012MNRAS.426.1905S,Pastorello2015MNRAS.449.1921P,Reguitti2022A&A...662L..10R}. As a further support, the CSM velocity (1900 \kms) measured for SN\,2022pda is consistent with that expected in WR winds (of the order of $10^3$ \kms), which is significantly faster than LBV winds \citep{Smith2017hsn..book..403S}. 

The broad spectral bumps at 4600-5200 \AA, 6400-6800~\AA\ regions in the spectra SN\,2022pda reveal a complex profile. This can be attributed to blends of H, He, and other emission lines, or to only H features coming from multiple emitting regions. We note that H spectral lines with a triple-peak profile, similar to that observed in SN\,2022pda spectra, were observed in the Type IIn SN 2010jp. For that object, \citet{Smith2012MNRAS.420.1135S} proposed a possible jet-like ejection of H-rich material. In this context, SN\,2022pda may have a similar asymmetric geometry as SN\,2010jp, with the CSM having a more He-rich composition. 

When the spectral lines of an interacting SN exhibit a composite profile, a plausible explanation is offered by invoking the interaction of the SN ejecta with an asymmetric CSM (for instance, with a disk-like structure or a torus). This scenario has been proposed for interacting transients showing two-humped light curves \citep[see][]{Gangopadhyay2025MNRAS.537.2898G,Dong2024ApJ...977..254D,Hiramatsu2024ApJ...964..181H,Gagliano2025ApJ...989..182G}. In the case of SN\,2022pda, a disk-like CSM formed through previous mass loss events. Hence, the first SN light-curve peak is driven by the strong interaction between the SN ejecta and the surrounding CSM in the equatorial region. After that, the second major interaction event occurred, but with a longer duration than the first one. This is likely due to the collision of the ejecta with a more expanded, lower-density CSM.

\subsection{Progenitor scenarios} \label{subsec:progenitor}

The general consensus on the origin of SNe Ibn is the terminal explosion of massive, H-poor WR stars within He-rich CSM \citep[e.g.,][]{Foley2007ApJ...657L.105F,Pastorello2007Natur.447..829P,Tominaga2008ApJ...687.1208T,Maeda2022ApJ...927...25M}, while LBVs are possible progenitors for SNe IIn and SN\,2009ip-like events \citep[e.g.,][]{Smith2010AJ....139.1451S, Foley2011ApJ...732...32F, Mauerhan2013MNRAS.430.1801M,Dessart2015MNRAS.449.4304D,DeLaRosa2017ApJ...850..133D,Boian2019A&A...621A.109B,Ransome2025ApJ...987...13R}. However, recent results suggest that they may have lower mass origins in interacting binaries \citep[see e.g.,][]{Anderson2012MNRAS.424.1372A,Kuncarayakti2018A&A...613A..35K,Sun2020MNRAS.491.6000S,Dessart2022A&A...658A.130D,Niu2026ApJ..1001...44N}. 
SN\,2022pda has somewhat hybrid properties between SNe Ibn and IIn, based on the observed photometric and spectroscopic properties, along with the light-curve modelling. Our findings suggest that the progenitor of SN\,2022pda was a late-type WR star with hydrogen (WNH), or even an Ofpe/WN9 \citep[][]{Conti1975MSRSL...9..193C,Walborn1977ApJ...215...53W,Walborn1982ApJ...256..452W,Bianchi2004ApJ...601..228B}. However, precursor activity and a highly asymmetric CSM  distribution (disk-like or bipolar) are  comfortably the outcomes of binary interactions, without necessarily invoking massive WR progenitors \cite[e.g.,][]{Sun2020MNRAS.491.6000S,Dessart2022A&A...658A.130D,Tsuna2024ApJ...966...30T}.

\subsection{Concluding remarks} \label{sec:conclusion} 
In this letter, we presented new data and the analysis of the photometric and spectroscopic properties of the interacting SN\,2022pda, which shows transitional features between those of a Type Ibn and a Type IIn SN. SN\,2022pda reveals the existence of a continuum in the properties, such as progenitor types (from WRs to LBVs), mass-loss mechanisms, CSM configuration, from SNe Ibn to SN IIn. The upcoming operation of 10-year Legacy Survey of Space and Time (LSST; \citealt{Ivezic2019ApJ...873..111I}) at the Vera C. Rubin Observatory\footnote{\url{https://www.lsst.org/}} will discover a large sample of transitional objects and reveal in detail the pre-SN photometric variability of SNe Ibn. Next-generation instruments such as the Chinese Space Station Telescope\footnote{\url{http://nao.cas.cn/csst/}} will be helpful for studying the progenitors of these and new species of ejecta-CSM interacting transients \citep{Fraser2021arXiv210807278F, Schulze2025Natur.644..634S}. Advanced models are required to accurately interpret the photometric and spectroscopic data, and to establish a clear picture of the nature of SN\,2022pda-like transients.

\vspace{5mm}
\facilities{Swift (UVOT), Swift (XRT), WISE, TRAPPIST, Moravian, LCO:fa01, LCO:fa03, LCO:fa04, LCO:fa05, LCO:fa06, LCO:fa07, LCO:fa11, LCO:fa12, LCO:fa15, LCO:fa16, LCO:fa19, LCO:fa20, LT (IO:O), LJT (YFOSC), NOT (ALFOSC), NOT (NOTCam), NTT (EFOSC2), NTT (SOFI), UH88 (SNIFS), SOAR (GHTS), VLT (X-Shooter), OGG2m (FLOYDS), COJ2m (en12).}

\software{astropy \citep{Astropy2013A&A...558A..33A,Astropy2018AJ....156..123A},  
          Matplotlib \citep{Hunter2007CSE.....9...90H}, 
          Numpy \citep{Harris2020Natur.585..357H},
          SciPy \citep{Virtanen2020NatMe..17..261V},
          HEASoft \citep[]{HEAsoft2014ascl.soft08004N},
          HOTPANTS \citep{Becker2015ascl.soft04004B},
          PESSTO pipeline \citep{Smartt2015A&A...579A..40S},
          SNIFS pipeline \citep{Tucker2022PASP..134l4502T},
          FLOYDS pipeline \citep{Valenti2014MNRAS.438L.101V},
          Source Extractor \citep{Bertin1996A&AS..117..393B}.}

\bibliography{Ibnref}{}
\bibliographystyle{aasjournal}

\appendix

\section{Ancillary information for photometric and spectroscopic data}
\label{Appendix:PhotSpec}

\subsection{Instrumental setups}\label{subsec:facilities}

\begin{table*}[ht!]
\caption{Log of photometric observations of SN\,2022pda.}
\label{table_setup}
\scalebox{0.67}{
\begin{tabular}{@{}lllll@{}}
\hline \hline
Code & Diameter&Telescope & Instrument & Site \\
                & $\mathrm{m}$&            &                   & \\
\hline
TRAPPIST       & 0.60 &TRAPPIST-S Telescope & FLI ProLine &       ESO La Silla Observatory, La Silla, Chile\\
Moravian  & 0.67/0.92 & Schmidt Telescope   & Moravian &  Osservatorio Astronomico di Asiago, Asiago, Italy\\
fa01$^*$       & 1.00 &  LCO (CPT site) &   Sinistro  & LCO node at South African Astronomical Observatory, Sutherland, South Africa \\
fa03$^*$       & 1.00 &  LCO (LSC site) &   Sinistro  & LCO node at Cerro Tololo Inter-American Observatory, Cerro Tololo, Chile   \\
fa04$^*$       & 1.00 &  LCO (LSC site) &   Sinistro  & LCO node at Cerro Tololo Inter-American Observatory, Cerro Tololo, Chile   \\
fa05$^*$       & 1.00 &  LCO (ELP site) &   Sinistro  & LCO node at McDonald Observatory, Texas, USA  \\
fa06$^*$       & 1.00 &  LCO (CPT site) &   Sinistro  & LCO node at South African Astronomical Observatory, Sutherland, South Africa \\
fa07$^*$       & 1.00 &  LCO (ELP site) &   Sinistro  & LCO node at McDonald Observatory, Texas, USA  \\
fa11$^*$       & 1.00 &  LCO (TFN site) &   Sinistro  & LCO node at Teide Observatory, Tenerife, Spain  \\
fa12$^*$       & 1.00 &  LCO (COJ site) &   Sinistro  & LCO node at Siding Spring Observatory, New South Wales, Australia  \\
fa15$^*$       & 1.00 &  LCO (LSC site) &   Sinistro  & LCO node at Cerro Tololo Inter-American Observatory, Cerro Tololo, Chile   \\
fa16$^*$       & 1.00 &  LCO (ELP site) &   Sinistro  & LCO node at McDonald Observatory, Texas, USA  \\
fa19$^*$       & 1.00 &  LCO (COJ site) &   Sinistro  & LCO node at Siding Spring Observatory, New South Wales, Australia  \\
fa20$^*$       & 1.00 &  LCO (TFN site) &   Sinistro  & LCO node at Teide Observatory, Tenerife, Spain  \\
IO:O           & 2.00 & Liverpool Telescope (LT)  & IO:O     &  Observatorio Roque de Los Muchachos, La Palma, Spain\\
LJT            & 2.40 & Lijiang 2.4\,m Telescope  & YFOSC      & Gaomeigu site, Lijiang Observatory (LJO), Yunnan, China\\
ALFOSC         & 2.56 & Nordic Optical Telescope (NOT) & ALFOSC  &   Observatorio Roque de Los Muchachos, La Palma, Spain\\
NOTCam         & 2.56 & Nordic Optical Telescope & NOTCam &  Observatorio Roque de Los Muchachos, La Palma, Spain\\
EFOSC2         & 3.58 & New Technology Telescope (NTT)& EFOSC2 & ESO La Silla Observatory, La Silla, Chile\\
SOFI           & 3.58 & New Technology Telescope & SOFI  & ESO La Silla Observatory, La Silla, Chile\\   
\hline \hline
\end{tabular}
}
\medskip
\\ 
\begin{flushleft}
$*$ These telescopes are distributed globally at different observatory sites and form a network of Las Cumbres Observatory (LCO) \citep{Brown2013PASP..125.1031B}. These photometric data come from the Global Supernova Project (GSP) and the ePESSTO+/LCO time, while the spectroscopic data are all from GSP.     
\end{flushleft}
\end{table*}

\begin{table*}[ht!] 
\caption{Log of spectroscopic observations of SN\,2022pda.}
\label{2022pdaSpecInfo}

 \scalebox{0.75}{
\setlength{\tabcolsep}{1.5pt}
\begin{tabular}{@{}cccccccc@{}}
\hline
Date & MJD & Phase$^a$ & Telescope+Instrument & Grism/Grating+Slit & Spectral range & Resolution & Exp. time \\ 
  &   & (days) &   &        & (\AA)    & (\AA)           & (s)           \\ 
\hline
20220925 & 59847.4 & $-$9.3   &  UH88+SNIFS  & B/R+? & 3400-9100  & R=1000 & 2700 \\
20220929 & 59851.1 & $-$5.6   &  SOAR+GHTS\_RED  &  400\_M1+1.0" & 3700-7000  &  6  &1800 \\
20220929 & 59851.1 & $-$5.6   &  VLT+X-Shooter & UVB/VIS/NIR+1.0" & 3100-24800  &R=5400/8900/5600 &{\fontsize{7}{12}\selectfont 2x1550/2x1500/2x300} \\
20220929 & 59851.2 & $-$5.5   &  OGG 2m+FLOYDS   & red/blu+2.0"  & 3500-10000  &  13   & 2699   \\
20220930 & 59852.1 & $-$4.6   &  NTT+EFOSC   & gr11/gr16+1.0"   & 3370-10010  & 14   &2400/2400 \\
20221001 & 59853.5 & $-$3.2   &  LJT+YFOSC & grism3+2.5"  & 3500-8760   &  26   & 2100  \\
20221008 & 59860.3 & $+$3.6   &  OGG 2m+FLOYDS & red/blu+2.0"  & 3500-10000  &  13   & 2700       \\
20221010 & 59862.9 & $+$6.2   &  NOT+ALFOSC  & gm4+1.0"        & 3540-9700  &  14&  2400      \\
20221020 & 59872.9 & $+$16.2  &  NOT+ALFOSC  & gm4+1.0"        & 3510-9700  &  14&  3000      \\
20221027 & 59879.4 & $+$22.7  &  COJ 2m+en12  & red/blu+2.0" & 3500-10000  & 13 &    3600    \\
20221029 & 59881.5 & $+$24.8  &  COJ 2m+en12  & red/blu+2.0" & 3500-10000  & 13 &    3600    \\
20221031 & 59883.1 & $+$26.4  &  NTT+EFOSC    & gr11/gr16+1.5" & 3350-9990 & 21 & 2x2100/2x2100 \\
20221104 & 59887.4 & $+$30.7  &  COJ 2m+en12  & red/blu+2.0"  &3500-10000   & 13 &    3600    \\
20221110 & 59893.5 & $+$36.8  &  COJ 2m+en12  & red/blu+2.0"  & 3500-10000  & 13 &    3600    \\
20221116 & 59899.5 & $+$42.8  &  COJ 2m+en12  & red/blu+2.0"  & 3500-10000  & 13 &    3600    \\
20221117 & 59900.8 & $+$44.1  &  NOT+ALFOSC  & gm4+1.3"       & 3720-9640   & 17 & 3000       \\
20221125 & 59908.8 & $+$52.1  &  NOT+ALFOSC  & gm4+1.1"       & 3500-9700   & 14 &  3600      \\
20221126 & 59909.4 & $+$52.7  &  COJ 2m+en12  & red/blu+2.0" & 3500-10000   & 13 &    3600    \\
20221202 & 59915.2 & $+$58.5  &  OGG 2m+en06& red/blu+2.0" & 3500-10000   & 13 &    3600    \\
20221215 & 59928.8 & $+$72.1  &  NOT+ALFOSC  & gm4+1.3"       & 3500-9650   & 17 &  3600      \\
\hline
\end{tabular}

 }
 
\medskip
$^a$Phases are relative to $r$-band maximum light (MJD = 59856.7). 
\end{table*}

\newpage

\subsection{Photometric data reduction} \label{subsec:reduphot}

All raw photometry images were first pre-reduced following the standard procedures in \textsc{iraf} \citep{Tody1986SPIE..627..733T, Tody1993ASPC...52..173T}, such as bias, overscan, and flat-field corrections. 
Multiple exposures were taken when the object was faint and then combined to increase the signal-to-noise ratio (S/N). The optical and NIR photometric data observed from ground-based telescopes were reduced using the dedicated pipeline {\sl ecsnoopy}\footnote{{\sl ecsnoopy} is a package for SN photometry using PSF fitting and/or template subtraction developed by E. Cappellaro. A package description can be found at \url{http://sngroup.oapd.inaf.it/snoopy.html}.}, which comprises several photometric packages, such as {\sc sextractor}~\citep[][]{Bertin1996A&AS..117..393B} for source extraction, {\sc daophot}~\citep[][]{Stetson1987PASP...99..191S} for magnitude measurement through point-spread function (PSF) fitting, and {\sc hotpants} \citep{Becker2015ascl.soft04004B} for image subtraction with PSF matching. The instrumental magnitudes of the SN are measured through the PSF-fitting technique, and calibrated against the catalogues of the \citet{Landolt1992AJ....104..340L}, Pan-STARRS, and the Two Micron All Sky Survey \citep[2MASS;][]{Skrutskie2006AJ....131.1163S}. 

We retrieved the Single Exposure frames by the NEOWISE-Reactivation mission \citep{Wright2010AJ....140.1868W, Mainzer2014ApJ...792...30M} - which scans the entire sky once every 6 months - from the IRSA Archive\footnote{https://irsa.ipac.caltech.edu/Missions/wise.html}. Then, we co-added them into a single image and performed the template-subtraction using images acquired on May 2014 (8 years before the explosion) as templates.
The WISE magnitudes were calibrated against the WISE All-Sky Data Release catalog \citep{Cutri2012wise.rept....1C}. \textit{Swift/UVOT} ultraviolet (UV) and optical photometry were retrieved from the NASA \textit{Swift} Data Archive\footnote{\url{https://heasarc.gsfc.nasa.gov/cgi-bin/W3Browse/swift.pl}}, and measured using the standard UVOT data analysis software {\tt HEASoft}\footnote{\url{https://heasarc.gsfc.nasa.gov/lheasoft/download.html}} \citep[version 6.19, ][]{HEAsoft2014ascl.soft08004N}, along with the standard calibration data. The source flux is measured within a circular aperture of $5^{''}$. Sky background was computed in a manually chosen region with a radius of $25^{''}$. 

The ZTF data were directly retrieved from the ZTF Forced Photometry Service \cite[][]{Masci2023arXiv230516279M}, while the ATLAS light curves \citep{Smith2020PASP..132h5002S}
were produced from the public Forced Photometry server\footnote{\url{https://fallingstar-data.com/forcedphot/}} \citep{Shingles2021TNSAN...7....1S}. Pan-STARRS (PS1+PS2) data were obtained through forced photometry on template-subtracted images following the similar method as in ATLAS dataset: in both cases we stacked all measurements on a given day into a single measurement 
\citep[for details of the Pan-STARRS transient survey data processing, see][]{Fulton2025MNRAS.542..541F}.

\subsection{Spectroscopic observation and data reduction} \label{subsec:spectdata}

The classification spectrum \citep{Fulton2022TNSAN.198....1F} was obtained with the University of Hawaii 2.2m telescope (UH88) plus the Supernova Integral Field Spectrograph (SNIFS) mounted on Mauna Kea, Hawaii, USA. The spectroscopic follow-up observations were carried out at the following facilities, including the X-Shooter spectrograph mounted on the 8.2m Very Large Telescope (VLT) at the ESO Paranal observatory, Chile; the Lijiang 2.4 m Telescope (LJT) equipped with the Yunnan Faint Object Spectrograph and Camera (YFOSC)~\citep{Wang2023RAA....23c5014W}, located at Lijiang Observatory, China; the 4.1m Southern Astrophysical Research Telescope (SOAR) with the Goodman High Throughput Spectrograph (GHTS) at Cerro Pach\'on in Chile; the Alhambra Faint Object Spectrograph and Camera (ALFOSC) on the NOT at the Observatorio del Roque de los Muchachos, La Palma, Spain, based on the Nordic-optical-telescope Unbiased Transient Survey 2 (NUTS2) project; the ESO Faint Object Spectrograph and Camera (EFOSC2) on the NTT at the La Silla Observatory, Chile, as part of the advanced Public ESO Spectroscopic Survey of Transient Objects (ePESSTO+) survey; the Las Cumbres Observatory (LCO; \citealt{Brown2013PASP..125.1031B}) with a global telescope network located at different sites: the FLOYDS (en06) spectrograph on the OGG 2m telescope at Haleakala Observatory, Maui, USA and the FLOYDS (en12) spectrograph on the COJ 2m telescope at Siding Spring Observatory, NSW, Australia, which are part of the Global Supernova Project (GSP). 

The spectra were reduced using routine {\sc iraf} tasks or with dedicated pipelines for specific instruments such as {\sc Foscgui}\footnote{Foscgui is a graphic user interface aimed at extracting SN spectroscopy and photometry obtained with FOSC-like instruments. It was developed by E. Cappellaro. A package description can be found at \url{http://sngroup.oapd.inaf.it/foscgui.html}.}, PESSTO pipeline \citep{Smartt2015A&A...579A..40S}, SNIFS pipeline \citep{Tucker2022PASP..134l4502T}~and FLOYDS pipeline\footnote{\url{https://lco.global/documentation/data/floyds-pipeline/}}. The different tools follow a standard manner: As a preliminary reduction, raw images were corrected for bias, overscan, trimming and flat fielding. Then, one-dimensional (1D) spectra were optimally extracted from the two-dimensional (2D) images. The wavelength calibration was performed using the arc-lamp spectra, while the flux of SN spectra was calibrated with the spectrophotometric standards. Both calibration images were taken during the same night as the SN observation. The flux calibrations of SN spectra were fine-tuned using the coeval broad-band photometry, and finally major telluric absorption lines (e.g. from O$_2$ and H$_2$O) were removed from the SN spectra using the spectra of standard stars.

\section*{Acknowledgments}
\begin{footnotesize}

We thank the anonymous referee for his/her insightful comments and suggestions that improved the paper.
This work is supported by the B-type Strategic Priority Program of the Chinese Academy of Sciences (Grant No. XDB1160202), the National Key R\&D Program of China with grants 2021YFA1600404, 2024YFA1611603, the National Natural Science Foundation of China (NSFC grants 12303054, 12173082, 12333008, 12288102, 12033003, and 11633002), the Yunnan Fundamental Research Projects (YFRP; grants 202401AU070063, 202501AV070012, 202501AS070078, and 202401BC070007), the Top-notch Young Talents Program of Yunnan Province, the Light of West China Program provided by the Chinese Academy of Sciences, and the International Centre of Supernovae (ICESUN), Yunnan Key Laboratory of Supernova Research (No. 202505AV340004). Y.-Z. Cai, AR, GV, and IS acknowledge financial support from the SOXS project (PI S. Campana). SC acknowledges funding from the Italian Space Agency (ASI/INAF Contract I/004/11/6. We thank
M. Fulton, A. Gkini, S. C. Williams, and M.-X. Huang for their assistance with observations and for sharing their data.

AP, AR, EC, GV, NER, IS, SB, and LT acknowledge support from the PRIN-INAF 2022, `Shedding light on the nature of gap transients: from the observations to the models'. AR also acknowledges financial support from the GRAWITA Large Program Grant (PI P. D'Avanzo).

M.F. acknowledges financial support of Taighde \'{E}ireann - Research Ireland under Grant number 24/FFP-P/12959. 

T.E.M.B. is funded by Horizon Europe ERC grant no. 101125877.

T.K. acknowledges support from the Research Council of Finland project 360274.

T.-W.C. acknowledges the Yushan Fellow Program by the Ministry of Education, Taiwan for the financial support (MOE-111-YSFMS-0008-001-P1).

SJB acknowledge their support by the European Research Council (ERC) under the European Union's Horizon Europe research and innovation programme (grant agreement No. 10104229 - TransPIre).

J. D. and R. P. S. acknowledge support by FCT for CENTRA through grant No. UID/PRR/00099/2025 (https://doi.org/10.54499/UID/PRR/00099/2025) and grant No. UID/00099/2025 (https://doi.org/10.54499/UID/00099/2025). J. D. acknowledges support by FCT under the PhD grant 2023.01333.BD, with DOI https://doi.org/10.54499/2023.01333.BD. R. P. S. acknowledges support by FCT under the PhD grant 2024.03599.BD.

L.G. acknowledges financial support from AGAUR, CSIC, MCIN and AEI 10.13039/501100011033 under projects PID2023-151307NB-I00, PIE 20215AT016, CEX2020-001058-M, ILINK23001, COOPB2304, and 2021-SGR-01270.

M.D. Stritzinger acknowledges support from the Independent Research Fund Denmark (IRFD; grant 10.46540/2032-00022B).

TP acknowledges the financial support from the Slovenian Research Agency (grants I0-0033, P1-0031, J1-8136, J1-2460 and Z1-1853). 

SMo is funded by Leverhulme Trust grant RPG-2023-240.

SZ is supported by the National Natural Science Foundation of China (NSFC, Grant No. 12473031), the Yunnan Fundamental Research Projects (Grant No. 202501AS070078)

XFW is also supported by the Tencent Xplorer Prize.

SJS and KWS acknowledge funding from STFC Grants ST/Y001605/1, ST/X001253/1, a Royal Society Research Professorship and the Hintze Family Charitable Foundation. 

This work was funded by ANID, Millennium Science Initiative, ICN12\_009

We acknowledge the support of the staffs of the various observatories at which data were obtained.

Based on observations collected at the European Organisation for Astronomical Research in the Southern Hemisphere, Chile, as part of ePESSTO+ (the advanced Public ESO Spectroscopic Survey for Transient Objects Survey - PI: Inserra). ePESSTO+ observations were obtained under ESO program IDs 108.220C and 111.24PR.

Based on observations made with the Nordic Optical Telescope, owned in collaboration by the University of Turku and Aarhus University, and operated jointly by Aarhus University, the University of Turku, and the University of Oslo, representing Denmark, Finland, and Norway, the University of Iceland, and Stockholm University at the Observatorio del Roque de los Muchachos, La Palma, Spain, of the Instituto de Astrofisica de Canarias.
Observations from the NOT were obtained through the NUTS2 collaboration which is supported in part by the Instrument Centre for Danish Astrophysics (IDA), and the Finnish Centre for Astronomy with ESO (FINCA) via Academy of Finland grant nr 306531. The data presented here were obtained in part with ALFOSC, which is provided by the Instituto de Astrofisica de Andalucia (IAA) under a joint agreement with the University of Copenhagen and NOTSA.

The Liverpool Telescope is operated on the island of La Palma by Liverpool John Moores University in the Spanish Observatorio del Roque de los Muchachos of the Instituto de Astrofisica de Canarias with financial support from the UK Science and Technology Facilities Council.

Funding for the LJT has been provided by Chinese Academy of Sciences and the People's Government of Yunnan Province. The LJT is jointly operated and administrated by Yunnan Observatories and Center for Astronomical Mega-Science, CAS.

Based on observations collected at Schmidt telescope (Asiago Mount Ekar, Italy) of the INAF -- Osservatorio Astronomico di Padova.

This work makes use of data from the Las Cumbres Observatory Network and the Global Supernova Project. The LCO team is supported by U.S. NSF grants AST-1911225 and AST-1911151.

We acknowledge the use of public data from the ATLAS, Pan-STARRS, ZTF, Swift and WISE data archive.

Y.-Z. Cai: This paper is dedicated to the memory of my grandfather, a centenarian who lived to the age of 103.  It is a poignant coincidence that this paper was accepted on the same day he was laid to rest.

\end{footnotesize}

\newpage

\end{document}